\documentstyle[preprint,version2,aps]{revtex}
\newcommand{\Nud}{\mbox{${{\cal N}^{\uparrow\downarrow}}(\vec{k})$}}
\newcommand{\R}{{\rm Re}}
\newcommand{\I}{{\rm Im}}
\newcommand{\Pu}{\mbox{$p_\uparrow$}}
\newcommand{\Pd}{\mbox{$p_\downarrow$}}
\newcommand{\Guu}[1]{\mbox{$G_{\uparrow \uparrow}^{#1}$}}

\newcommand{\Gdd}[1]{\mbox{$G_{\downarrow \downarrow}^{#1}$}}
\newcommand{\Nu}[1]{\mbox{$n^\uparrow_{\vec{#1}}$}}
\newcommand{\Nd}[1]{\mbox{$n^\downarrow_{\vec{#1}}$}}
\newcommand{\Nt}[1]
{\mbox{$n_{\uparrow\downarrow}(t,\vec{r},\vec{#1})$}}

\newcommand{\Gt}[1]{\mbox{$\tilde{G}^{#1}_{\uparrow\downarrow}$}}
\newcommand{\St}[1]{\mbox{$\tilde{\Sigma}^{#1}_{\uparrow\downarrow}$}}
\newcommand{\Suu}[1]{\mbox{$\Sigma^{#1}_{\uparrow\uparrow}$}}
\newcommand{\Sdd}[1]{\mbox{$\Sigma^{#1}_{\downarrow\downarrow}$}}
\newcommand{\Eu}[1]{\mbox{$\epsilon_\uparrow(\vec{#1})$}}
\newcommand{\Ed}[1]{\mbox{$\epsilon_\downarrow(\vec{#1})$}}

\begin{document}
\draft
\begin{quote}
\raggedleft cond-mat/9707020
\\{\em J. Low Temp. Phys.}, vol. 112, in press (1998)
\end{quote}
\begin{title}
\centering
Transverse Spin Diffusion~~~~~~~~~~~~~~~~~~~~ \\
in a Dilute Spin-Polarized Degenerate Fermi Gas
\end{title}

\author{Denis I. Golosov\footnotemark[1] and Andrei
E. Ruckenstein\footnotemark[2] }
\begin{instit}
Department of Physics, Rutgers University, Piscataway, NJ 08855-0849,
U.S.A.
\end{instit}
\footnotetext[1]{Present address: Argonne National Laboratory, Materials 
Science Division, 9700 S. Cass Ave., Argonne, IL 60439, U. S. A.
E-mail: golosov@franck.uchicago.edu}
\footnotetext[2]{Also at: Insitut f\"{u}r Theorie der Kondensierten
Materie, Universit\"{a}t Karlsruhe, 
76132 Karlsruhe, Germany. E-mail: andreir@physics.rutgers.edu} 
\begin{abstract}
We re-examine the calculation of the transverse spin-diffusion coefficient
in a dilute degenerate spin-polarized Fermi gas, for the case of
$s$-wave scattering. The special feature of this limit is that the dependence
of the spin diffusion coefficient on temperature
and field can be calculated explicitly with no further approximations.
This exact solution uncovers a novel intermediate behaviour between
the high field spin-rotation dominated regime in which 
$D_{\bot} \propto H^{-2}$, $D_{\parallel} \propto T^{-2}$, and the
low-field isotropic, collision dominated regime with 
$D_{\bot} = D_{\parallel} \propto T^{-2}$.  In this intermediate
regime, 
$D_{\bot ,\parallel} \propto T^{-2}$ but 
$D_{\bot} \neq D_{\parallel}$. 
We emphasize that the 
low-field crossover cannot be described within the
relaxation time approximation.
We also present an analytical calculation of the self-energy
in the s-wave approximation for a dilute spin-polarized
Fermi gas, at zero temperature. This emphasizes the
failure of the conventional Fermi-liquid phase space arguments
for processes involving spin flips.
We close by reviewing the evidence for the existence of the intermediate
regime in experiments 
on weakly spin-polarized
$^3{\rm  He}$ and $^3{\rm He} - ^4{\rm He}$ mixtures.
\end{abstract}

\section{Introduction}
\label{sec:introtau}

The unusual features of spin dynamics in spin-polarized quantum
systems have been intensively studied since the pioneering
paper of Leggett and Rice~\cite{Leggett} on spin
diffusion in normal liquid $^3{\rm He}$.
The main effect arises from the observation that the presence of a
molecular field (induced by the applied magnetic field) leads to
an additional precession of the spin-current which in
steady state acquires a component perpendicular to the
magnetization gradient; through the continuity equation this results
in an anomalous reactive component (damped spin-wave) to spin transport.
This ``spin-rotation" effect is also present in the case of
spin-polarized Boltzmann gases~\cite{gases}.

From a microscopic point of view,
it was expected that a number of
qualitatively novel phenomena might arise in the case of
sufficiently high polarizations~\cite{Metamagnetism}.
A natural suggestion, made by Meyerovich~\cite{Meyer85}, was that
low-temperature spin-diffusion becomes highly anisotropic for finite
polarizations.
More precisely,
processes involving
spin-flips make use of the phase-space volume between the
two distinct
Fermi surfaces (for ``up" and ``down" spins) leading to
a {\em finite} scattering rate in the limit of $T\rightarrow 0$. This is
in contrast with processes involving scattering in the vicinity
of each of the Fermi surfaces which are subject to the
phase-space restrictions of unpolarized Fermi liquids and are thus
characterized by the
conventional Fermi-liquid behaviour of scattering rates,
$\propto T^2$~\cite{AbrikosovFL}.
Meyerovich's suggestion was recently supported
by measurements of
the transverse spin-diffusion coefficient in weakly polarized liquid
$^3{\rm He}$~\cite{Candela}.
Theoretically, $D_\bot$ was calculated in the dilute gas limit
at $T=0$~\cite{Mullin89,KarenPR}; while for $T\neq 0$ the only
available estimates are based on a variational solution of
the Boltzmann equation ~\cite{Mullin89,Mullin92}, or on a phenomenological
Mathiessen-type rule \cite{Karen3}.

In this paper we present the analytical calculation of the finite
temperature behaviour of $D_\bot$ for a dilute  Fermi gas
in the s-wave
approximation, by solving the appropriate kinetic equations exactly.
While at $H\stackrel {>}{\sim}T$ our results for $D_\bot$ agree with
those of previous work \cite{Mullin89,Mullin92}, in the opposite case
of $T \gg H$ the exact solution of kinetic equation results in a
qualitatively new behaviour of $D_\bot$. We show, that the crossover
between the high-field and low-field behaviours is described by {\em
two} dimensionless parameters rather than only one as implied in
Refs. \cite{Mullin89,Mullin92,Karen3}:
The first is the ratio $H/T$, which determines whether the phase
space that is available for interparticle scattering is due to 
the field-induced splitting of the Fermi spheres ($H \gg T$) or to the
temperature smearing of the Fermi distribution function ($T \gg H$).
The diffusion coefficient is, roughly speaking, inversely proportional to the
square of the available  phase space volume; and hence in the two limits
we obtain  $D_\bot \propto H^{-2}$ and  $D_\bot
\propto T^{-2}$, respectively. The second parameter,
\begin{equation}
\zeta=(H/T)(ap_F)^{-1}(\epsilon_F/T),
\label{eq:zeta}
\end{equation}
($a$ is the $s$-wave scattering length and $\epsilon_F$ and $p_F$
are Fermi energy and momentum in the absence of magnetic field)
is the ratio of two frequency scales appearing in the kinetic equations
for transverse spin diffusion, $\zeta = \omega _1 /\omega _2$.
The scale $\omega _2$ is determined by the thermal term in the transverse 
 relaxation 
rate, $\omega_2 \sim 1/\tau_T \sim m a^2 T^2$, where $m$ is particle mass;
while $\omega _1 \sim ap_F H$ is the molecular field
contribution to the precession frequency
of the transverse spin current.
(Its origin
is the spin-dependent exchange scattering
amplitude of two particles interacting 
through a spin-independent interparticle
potential~\cite{Leggett}). 
Thus, the value of $\zeta$
measures the number of oscillations of the spin
current during a collision time.
For $\zeta \gg 1$ the ``spin-rotation" effect strongly affects the spin
diffusion process, which remains anisotropic even for $H\ll T$: in this case, 
we obtain $D_\bot \approx 0.89 D_\parallel$. The crossover to isotropic
behaviour occurs for $\zeta \sim 1$, so that  $D_\bot
\approx D_\parallel$ only once $\zeta \ll 1$.

We note that the crossover between isotropic and anisotropic
behaviours of $D_{\bot,\parallel}$ at $\zeta \sim 1$ has already been
discussed  previously \cite{Pal,Bedell}.
However, the qualitative changes in the collision integral which
occur at $H \sim T$ were overlooked. Similarly, the 
subtleties that arise in the solution of Boltzmann equation at $\zeta
\sim 1$ have not been noticed in
Refs. \cite{Mullin89,Mullin92,Karen3}, which however provide the
correct results for $\zeta \gg 1$. In the present paper, we will give
the complete description of field and temperature dependence of
$D_\bot$ for a dilute gas (within the $s$-wave approximation for the
two-body scattering) which is valid for {\em any} value of $H$ provided that 
the gas remains degenerate for both values of spin projection,
\begin{equation}
T \ll \frac{p_{\uparrow, \downarrow}^2}{2m}\,,
\end{equation}
where 
$p_{\uparrow, \downarrow}$ are the
Fermi momenta for spin-up and spin-down particles:
\begin{equation}
\frac{\Pu^2}{2m}-\frac{\Pd^2}{2m}=H\,,\,\,\,\,\,
\frac{\Pu^3+\Pd^3}{6 \pi^3}=N_\uparrow+N_\downarrow=N\,,\,\,\,\,
\frac{\Pu^3-\Pd^3}{12\pi^3}=M_\parallel\,,
\end{equation}
$N_{\uparrow,\downarrow}$ are the
numbers of spin-up and
spin-down particles per unit volume, $M_\parallel$ is the 
longitudinal magnetization per
unit volume induced by the field $H$, and
$N$ is the density of the gas. A short account of the present work
was already published in Ref. \cite{We}.

We emphasize that throughout this paper we concentrate on the hydrodynamic 
regime defined by {\em finite} longitudinal and transverse
relaxation times, $\tau_{\bot, \parallel}$, 
which, in principle, satisfy the condition $D_{\bot , \parallel} k^2
\tau_{\bot, \parallel} \ll 1$,   
where $k \sim 1/L \rightarrow 0$ ($L$ is the size of the system)
is the characteristic wavevector of
magnetization fluctuations.
The $T\rightarrow 0$ limit
of our calculations
leads to different physics from that of the strict $T=0$,
collisionless regime discussed by I. A. Fomin \cite{Fomin}.
Indeed, one should interpret our $T=0$ results as describing
fluctuations with sufficiently long wavelength, and at
low {\em but finite} temperatures, $T\ll H$,
such that
the hydrodynamic condition is satisfied (note that $D_\parallel$ and
$\tau_\parallel$ diverge at $T \rightarrow 0$). 
The interesting issue of the validity of the linearized Boltzmann 
equation approach in the context
of spin-polarized Fermi gases at low temperatures, raised in
Refs. \cite{Fomin,Thesis}, is left for future investigations.
Here, we limit ourselves to the linearized equations, which should be valid
even for $T\ll H$ as long as the
external probe used to drive the system
out of equilibrium is weak enough and 
sufficiently slowly varying in space.

The plan of the paper is as follows:
In Section \ref{sec:zeroT} we discuss the zero temperature limit of 
the kinetic equations derived more generally in Appendix \ref{app:kinetics}. 
To get a better feeling for the physical content of these equations we
begin with an analysis of the relaxation time approximation. We then
solve the $T \rightarrow 0$ equations exactly and recover the variational
result of Ref. \cite{Mullin89}.
This is followed in Section \ref{sec:smallpol} with the discussion of
the finite temperature equations in the limit of small polarization,
$H \ll \epsilon_F$, and the resulting exact solution for the diffusion
coefficient. 
Since the results of Section \ref{sec:zeroT}
are valid everywhere in the region $H \gg T$  our expressions are thus covering
the entire range of field and temperature values in the degenerate    
regime, $T \ll \epsilon _F$. 
This analysis uncovers the two crossovers already discussed above.
In Sect. \ref{sec:discu},
our results are compared with the available
experimental data in
$\,^{3}\rm{He} $--$\,^{4}\rm{He}$ mixtures
\cite{Owers95,Candela91,Nunes,Owers} and in 
$\,^{3}\rm{He}$ \cite{Candela}.
The derivation of the self-energies (which are used as an input into
the Boltzmann equations) is relegated to Appendix \ref{app:self}.
The calculations are instructive as they uncover the simple
phase-space origin of the finite value of $D_{\bot}$ at $T\rightarrow 0$.

Throughout  we use the  units with $k_B=\mu_0=\hbar=1$, where $k_B$ is
Boltzmann constant and $\mu_0/2$ is the magnetic moment of a particle;
also we take the volume of the system, $V=1$.

\section{Transverse Spin Diffusion at
${\boldmath T \rightarrow 0}$ and the Transverse Relaxation Time} 
\label{sec:zeroT}

We begin with the kinetic equations for the transverse (off-diagonal
in spin space) component of the density matrix in the rotating frame,
\Nt{p}
(see Appendix \ref{app:kinetics}):
\begin{eqnarray}
&&\left(\frac{\partial}{\partial t}+\frac{\vec{p}}{m}
\frac{\partial}{\partial \vec{r}} \right) \Nt{p} +
\frac{4\pi a}{m}i\int n_{\uparrow
\downarrow}(t,\vec{r},\vec{p}^{\,\prime})
\frac{d^3p^{\prime}}{(2\pi)^3}(\Nu{p}-\Nd{p})- 
\frac{4\pi a}{m}i \times \nonumber \\
&&\times(N_\uparrow-N_\downarrow)\Nt{p}
=-i\left\{\left[\left(\Suu{(2)}(\Eu{p},\vec{p})\right)^\ast-
\left(\Sdd{(2)}(\Ed{p},\vec{p})\right)^\ast\right]\Nt{p}+ \right.
\nonumber \\
&&+\left. \St{(2)++}\left(t,\vec{r};\frac{p^2}{2m}-\mu,\vec{p}\right)
(\Nu{p}-\Nd{p})\right\}\,
\label{eq:kin}
\end{eqnarray}
where $\mu$ is the chemical potential and
$\Sigma^{(2)}_{\uparrow \uparrow , \downarrow
\downarrow}$ denote the contributions in
Eqns. (\ref{eq:sigup}--\ref{eq:sigdowni}) quadratic in the scattering 
length $a$; similarly, the transverse self energy, 
$\St{(2)++}\left(t,\vec{r};\frac{p^2}{2m}-\mu,\vec{p}\right)$,
is given by Eqn. (\ref{eq:selftransverse}). In the $T \rightarrow 0$
limit, the Fermi distribution functions for spin-up and spin-down particles,
$n^\uparrow_{\vec{p}}$ and $n^\downarrow_{\vec{p}}$, can be
approximated by $\theta(p_{\uparrow,\downarrow}-p)$.

It is easy to verify that the r.\ h.\ s. of Eqn. (\ref{eq:kin}) does
not mix different Legendre components of the quantity \Nt{p} as a
function of momentum. Therefore, we will search for the solutions of
the form
\begin{equation}
\Nt{p}=g(t,\vec{r},p)+f(t,\vec{r},p)\cos \psi \,,
\label{eq:legendre}
\end{equation}
where $\psi$ is the angle formed by vector $\vec{p}$ and the $x$ axis.
It turns out that in the $s$-wave approximation the higher Legendre
harmonics do not contribute to \Nt{p} (see below).
Eqn. (\ref{eq:legendre}) implies that, in the rotating frame, the transverse
magnetization and spin current 
are given, respectively, by 
\begin{equation}
M_\bot^-(\vec{r},t) \equiv M_\bot^x(\vec{r},t) - i
M_\bot^y(\vec{r},t) = \frac{1}{2} \int  \Nt{p}
\frac{d^3 p}{(2\pi)^3} = \frac{1}{4 \pi^2 } \int_0^\infty g(t, \vec{r},
p) p^2 dp\,
\label{eq:defM}
\end{equation}
and
\begin{eqnarray}
J_\bot^-(\vec{r},t) &\equiv&  J_\bot^x(\vec{r},t) - i
J_\bot^y(\vec{r},t) = \frac{1}{2m} \int p \Nt{p}\cos \psi
\frac{d^3 p}{(2\pi)^3} = \nonumber \\
&=&\frac{1}{12 \pi^2 m} \int_0^\infty f(t, \vec{r},
p) p^3 dp\,,
\label{eq:defJ}
\end{eqnarray}
where the indices $x$ and $y$
refer to  spin space. 

To find the transverse spin diffusion coefficient $D_\bot$
one has to calculate the value of the spin current $J_\bot$ induced by
a slow spatial variation of the transverse
magnetization. We will assume, without loss of generality, that
$\vec{\nabla} M_\bot^-$ is uniform and is directed along the $x$  axis:
\begin{equation}
g(t,\vec{r},p)=(G_1+G_2 \cdot x)g_0(t,p)\,,\,\,\,\,\frac{\partial}
{\partial x}M_\bot^-= \frac{G_2} {4 \pi^2} \int_0^\infty g_0(t,p) p^2 dp\,,
\label{eq:ansatzg}
\end{equation}
where $\mid G_1+G_2 \cdot x \mid \ll 1$ for any $x$ inside the sample.

The value of $D_\bot$ should then be found from the macroscopic
constitutive relation \cite{Leggett,Helium3}, which can be written as
\begin{equation}
J^-_\bot= - \frac {D_\bot}{1-i \xi} \frac{\partial}{\partial x}
M_\bot^-\,,
\label{eq:difmacro}
\end{equation}
where the quantity $\xi$ is referred to as the {\em spin-rotation
parameter}.  
The non-zero imaginary part of the denominator in (\ref{eq:difmacro})
reflects the fact that, due to the spin-rotation effect, the spin current
is not parallel  to the driving magnetization gradient.

Substituting Eqns. (\ref{eq:legendre}) and (\ref{eq:ansatzg}) in the
Boltzmann equation (\ref{eq:kin}) and separating the Legendre
harmonics, we obtain two coupled kinetic equations:
\begin{eqnarray}
\frac{\partial}{\partial t} g_0(t,p)&+&\frac{1}{G_1+G_2\cdot x} \cdot
\frac{p}{3m}\frac{\partial }{\partial 
x}f(t,\vec{r},p) 
 + \frac {2a}{\pi m} i\int_0^\infty g_0(t,q) q^2 dq (\Nu{p} - \Nd{p})-
\nonumber \\
&-& \frac{4 \pi a}{m}i(N_\uparrow- N_\downarrow)g_0(t,p) 
= -I_{rel}[g_0]-iI_{sr}[g_0]
\label{eq:legen0}
\end{eqnarray}
and
\begin{equation}
\frac{\partial}{\partial t}f(t,\vec{r},p) + \frac{p}{m} G_2 g_0(t,p)
- \frac{4 \pi a}{m}i(N_\uparrow- N_\downarrow)f(t,\vec{r},p)
=-I_{rel1}[f]-iI_{sr1}[f]\,. 
\label{eq:legen1}
\end{equation}
where the linear operators representing the ``relaxation'' and
``spin-rotation'' contributions to the collision integral are defined (for an
arbitrary function $W(\vec{p})$) as
\begin{eqnarray}
I_{rel}[W(\vec{p})]&=&\left[\I\Suu{(2)}(\Eu{p},\vec{p})-
\I\Sdd{(2)}(\Ed{p},\vec{p})\right] W(\vec{p})-
\nonumber \\
&-&\pi\frac{(8\pi a)^2}{m}
(n^\uparrow_{\vec{p}}-n^\downarrow_{\vec{p}})
\int
\left[1-n^\uparrow_{\vec{s}+\vec{k}}-n^\downarrow_{\vec{s}-\vec{k}}+
2n^\uparrow_{\vec{s}+\vec{k}} 
n^\downarrow_{\vec{s}-\vec{k}}\,
\right] \times \label{eq:IrelT0} \\
&\times&\delta\left((\vec{p}-\vec{p}^{\,\prime})^2-4k^2\right)
W(\vec{p}^{\,\prime})
\frac{d^3k\,d^3p^\prime}{(2\pi)^6}\,,
\nonumber \\
I_{sr}[W(\vec{p})]&=&\left[\R\Suu{(2)}(\Eu{p},\vec{p})-
\R\Sdd{(2)}(\Ed{p},\vec{p})\right] W(\vec{p})-
\nonumber \\
&-&\frac{(8\pi a)^2}{m}
(n^\uparrow_{\vec{p}}-n^\downarrow_{\vec{p}})
\int {\cal P}
\frac{n^\uparrow_{\vec{s}+\vec{k}}+
n^\downarrow_{\vec{s}-\vec{k}}}{(\vec{p}-\vec{p}^{\,\prime})^2-4k^2} 
W(\vec{p}^{\,\prime}) 
\frac{d^3k\,d^3p^\prime}{(2\pi)^6}\,,
\label{eq:IsrT0}
\end{eqnarray}
where $\vec{s}=(\vec{p}+\vec{p}^{\,\prime})/2$, and
\begin{equation}
I_{rel1}[W(\vec{p})]=\frac{I_{rel}[W(\vec{p})\cos
\psi]}{\cos\psi}\,,\,\,\,\,\,\, 
I_{sr1}[W(\vec{p})]=\frac{I_{sr}[W(\vec{p})\cos\psi]}{\cos\psi}\,.
\label{eq:Ilegen1}
\end{equation}

The simplest way of understanding the physical meaning of 
the kinetic equations (\ref{eq:legen0}--\ref{eq:legen1})
is to appeal to the relaxation time approximation
for the collision integrals, Eq. (\ref{eq:Ilegen1}),
\begin{equation}
I_{rel1}[W(p)] = \frac{1}{\tau_\bot} W(p)\,,\,\,\,\,\,\,
I_{sr1}[W(p)] = -\Omega^{(2)}W(p)\,,
\end{equation}
where $\tau_\bot$ is the transverse relaxation
time, while $\Omega^{(2)}$ can be identified as a second-order
correction to the precession frequency of the spin-current (see below).
Upon integration, Equations (\ref{eq:legen0}--\ref{eq:legen1})
then read \cite{Meyer85} (see also Appendix \ref{app:kinetics}):
\begin{eqnarray}
\frac{\partial}{\partial t} M_\bot^- + \frac{\partial}{\partial x}
J^-_\bot &=& 0\,, 
\label{eq:macromagT0}\\
\frac{\partial}{\partial t}
J^-_\bot+\frac{p_\uparrow^2}{5m^2}\frac{1-d^5}{1-d^3}
\frac{\partial}{\partial x} M^-_\bot-
i(\Omega^{(1)}+\Omega^{(2)})J_\bot^- &=& -\frac{1}{\tau_\bot}J^-_\bot\,, 
\label{eq:macrocurT0}
\end{eqnarray}
where $d=\Pd/\Pu$ and 
\begin{equation}
\Omega^{(1)}= \frac{4 \pi a}{m}(N_\uparrow-N_\downarrow)=\frac{8\pi
a}{m} M_\parallel\,.
\label{eq:omega1T0}
\end{equation}
The main feature of Eqn. (\ref{eq:macrocurT0}) is
the precession of the spin current around the molecular field
$(\Omega^{(1)}+\Omega^{(2)})$ originating from the exchange
contributions to the single-particle self energies. This does not
appear in the equation for magnetization (Eqn. (\ref{eq:macromagT0}))
which only precesses at the bare Larmor frequency.
In the simple case taken here, $\partial M^-_\bot
/ \partial t = \partial J^-_\bot /\partial x =0$ and the solution to
Eqn. (\ref{eq:macrocurT0}) reads:
\begin{equation}
J_\bot^-(t)=- \frac {D_\bot}{1-i \xi} \frac{\partial}{\partial x}
M_\bot^-+ J^-_{\bot tr} (0) \exp\left[ i(\Omega^{(1)}+\Omega^{(2)})t -
\frac {t}{\tau_\bot} \right]\,.
\label{eq:macrocurT0solution}
\end{equation}
Thus, the steady state current takes the form (\ref{eq:difmacro}) with 
\begin{equation}
D_\bot \approx \frac{p_\uparrow^2}{5m^2} \frac{1-d^5}
{1-d^3} \tau_\bot\,,\,\,\,\,\xi=(\Omega^{(1)}+\Omega^{(2)})
\tau_\bot\,.
\label{eq:tauapprox}
\end{equation}
($ J^-_{\bot tr} (0)$ represents the initial value
of transient spin 
current which in our case is independent of $x$.)

Having used the relaxation time approximation to illustrate the
physical content of our kinetic equations
(\ref{eq:legen0}--\ref{eq:legen1}) for the transverse 
density matrix we now proceed with the exact solution of these equations.

For the purpose of computing the spin diffusion coefficient it is
sufficient to consider the steady-state
solution (\ref{eq:legendre}) with a time- and coordinate-independent
functions $f(t,\vec{r},p) \equiv f(p)$ and $g_0(t,p)
\equiv g_0(p)$.  In this case, the first two terms
on the l.\ h.\ s. of Eqn. (\ref{eq:legen0}) vanish along with the first
term on the l.\ h.\ s. of (\ref{eq:legen1})
\footnote{The only coupling between the first and second Legendre components
would involve $\partial f/\partial x$ which vanishes in our case in the
steady state. This justifies the form assumed in Eqn. (\ref{eq:legendre}).}.
With the help of Eqns. (\ref{eq:sigup}--\ref{eq:sigdown}) one can
easily check that Eqn. (\ref{eq:legen0}) is then solved exactly by
\begin{equation}
g_0(p)=n^\uparrow_{\vec{p}}-n^\downarrow_{\vec{p}}\,.
\label{eq:mageigen}
\end{equation}

To proceed with the solution of the Boltzmann equation we note that,
already on symmetry grounds, we expect
\begin{equation}
I_{rel1}[f] \sim m a^2 H^2 f\,,\,\,\,\,I_{sr1}[f] \sim  (N^{1/3}a)^2 H f\,,
\end{equation}
which also follows from explicit integration: after
a rather cumbersome calculation we find that for any function
$W(p)$ that vanishes outside the region $p_\downarrow<p < p_\uparrow$
(as required by (\ref{eq:domain1})),
\begin{equation}
\int_0^\infty I_{rel1}[W(p)]p^3 dp = \frac{a^2\Pu^4}{3\pi m}
\int_d^1 F(x,d)W(x\Pu)dx\,,
\label{eq:intIrel1T0}
\end{equation}
where  
\begin{eqnarray}
F(x,d)&=&\frac{16}{35}x^7-2x^4d^3-\frac{1}{5}x^2d^5+x^2d^3+
\frac{3}{35}d^7+ \frac{1}{5}d^5+\nonumber\\
&&+\frac{2}{35}\theta(x^2+d^2-1)\cdot(8x^2+d^2-1)
(x^2+d^2-1)^{5/2}\,.
\label{eq:F}
\end{eqnarray}

Thus, the two terms on the r.\ h.\ s. of Eqn. (\ref{eq:legen1}) are 
small compared to the third term on the l.\ h.\ s. which is of order
$N^{1/3}a H f$. Therefore, Eqn. (\ref{eq:legen1}) may  be
solved iteratively by substituting
\begin{equation}
f(p) \approx -i \frac{G_2 p}
{4 \pi a (N_\uparrow-N_\downarrow)}
(n^\uparrow_{\vec{p}}-n^\downarrow_{\vec{p}})
\label{eq:leadingT0} 
\end{equation}
on the r.\ h.\ s. . Upon integration, Eqn. (\ref{eq:difmacro}) then leads to:
\begin{eqnarray}
D_\bot&\approx&\frac{1}{25 m^2} \frac{(p_\uparrow^5-p_\downarrow^5)^2}
{p_\uparrow^3-p_\downarrow^3} \cdot \frac {1} {\int_0^\infty
I_{rel1}[pg_0(p)]p^3 dp}\,,
\label{eq:difgenericT0} \\
\xi&\approx& \frac{16 \pi^2 a} {15 m}
(p_\uparrow^5-p_\downarrow^5)(p_\uparrow^3-p_\downarrow^3)\cdot\frac {1}
{\int_0^\infty I_{rel1}[pg_0(p)]p^3 dp}- \nonumber \\
& -& \frac{\int_0^\infty
I_{sr1}[pg_0(p)]p^3 dp} {\int_0^\infty I_{rel1}[pg_0(p)]p^3 dp}\,.
\label{eq:xigenericT0}
\end{eqnarray}

In turn, Eqns. (\ref{eq:mageigen}),
(\ref{eq:intIrel1T0}--\ref{eq:F}), and 
(\ref{eq:difgenericT0}--\ref{eq:xigenericT0}) can now be used to
extract the final expressions
for $D_\bot$ and $\xi$,
exact 
in leading order in $N^{1/3}a$:
\begin{equation}
D_\bot \approx \frac{p_\uparrow^2}{5m^2} \frac{1-d^5}
{1-d^3} \tau_D\,\,,\,\,\,\,\,\,\xi \approx \frac{4 \pi
a}{m}(N_\uparrow -N_\downarrow)\tau_D\,,
\label{eq:difT0}
\end{equation}
where the {\em diffusion time} $\tau_D$ is given by
\begin{eqnarray}
\frac{1}{\tau_D}& =& \frac{16 a^2}{189\pi m}\cdot \frac
{p_\uparrow^4}{1-d^5} \left\{
1-\frac{21}{64}d^3(1-d^2)(5+2d^2+17d^4)-\right.
\nonumber \\
&-&\left. \frac{1}{8} \theta(2d^2-1)\cdot(1+7d^2)(2d^2-1)^{7/2}
\right\}\,.
\label{eq:diftimeT0}
\end{eqnarray}
This proves that the result originally reported as {\em variational}
in Ref. \cite{Mullin89} is in fact exact in the appropriate limit 
(we note that this result has been substantiated earlier through
the numerical evaluation of the expressions in Ref. \cite{KarenPR}). 
Equations (\ref{eq:difT0}--\ref{eq:diftimeT0}) show that the
transverse spin diffusion coefficient remains {\em finite} at $T
\rightarrow 0$,
as was already discussed above. In particular,
in the limit of low polarizations,  when
$1-d\approx m H p_F^{-2} \ll 1$, we find 
\begin{equation}
\tau_D (H)=\frac{9}{8}\frac{\pi}{ma^2}\, H^{-2}\,.\,\,\,
\label{eq:bounds}
\end{equation}
The detailed results for the spin diffusion time, $\tau _{D}$, are
plotted in Fig. \ref{fig:bounds} (dashed line) together with
variational bounds, $\tau_\bot^\pm$,
on the relaxation time, $\tau _\bot$, which appears in the relaxation time
approximation \cite{bounds}.

\section{Transverse Spin Diffusion at Finite Temperatures and Small
Polarizations} 
\label{sec:smallpol}

In this section, we 
extend our results to the case of finite temperatures in weakly
polarized limit, within the degenerate regime, $T \ll \epsilon_F$.
The appropriate Boltzmann equation is derived in the Appendix 
\ref{app:kinetics} (see Eqn. (\ref{eq:Diffkin})).
The only difference from the $T \rightarrow 0$ case of
Eqns. (\ref{eq:legendre} -- 
\ref{eq:Ilegen1}) is in the form of collision terms,
\begin{equation}
I_{rel}[W(\vec{p})]=W(\vec{p})\,  \int A(\vec{p},\vec{p}^{\,\prime})
\frac{ d^3 p^\prime}{(2 \pi)^3}- \int A(\vec{p}^{\,\prime},\vec{p})
W(\vec{p}^{\,\prime})
\frac{ d^3 p^\prime}{(2 \pi)^3}\,\,,
\label{eq:reltot}
\end{equation}
\begin{equation}
I_{sr}[W(\vec{p})]= 
W(\vec{p}) \int 
B(\vec{p},\vec{p}^{\,\prime}) (n^\uparrow_{\vec{p}^{\,\prime}}-
n^\downarrow_{\vec{p}^{\,\prime}}) \frac {d^3 p^\prime}{(2 \pi)^3} -
 (n^\uparrow_{\vec{p}}-
n^\downarrow_{\vec{p}}) \int B(\vec{p},\vec{p}^{\,\prime})
W(\vec{p}^{\,\prime}) \frac {d^3 p^\prime}{(2
\pi)^3} \,\,.
\label{eq:spinrotcoll}
\end{equation}
Here
\begin{equation}
A(\vec{p},\vec{p}^{\,\prime}) \approx \frac{ \pi a^2}{p}
\left(\frac{(p^2+{p^\prime}^2)}{m}- 4\mu\right)\cdot\left(
n^\uparrow_{\vec{p}^{\,\prime}}+
n^\downarrow_{\vec{p}^{\,\prime}}-\frac{1}{1-{\rm exp} \left[
\frac{p^2+{p^\prime}^2}{2mT}-\frac{2\mu}{T}
\right]} \right)\,,
\label{eq:aaproxT}
\end{equation}
and
\begin{equation}
B(\vec{p},\vec{p}^{\,\prime}) \approx \frac{2a^2}{m} \left( -4p_F + |
\vec{p} - \vec{p}^{\,\prime}| \ln \left|\frac{|\vec{p} -
\vec{p}^{\,\prime}| + 2p_F}{|\vec{p} -
\vec{p}^{\,\prime}| - 2p_F}\right|\right)\,
\label{eq:baproxT}
\end{equation}
(here $p_F=(2m\epsilon_F)^{1/2}$ is the Fermi momentum of
the non-polarized system).
Once the kinetic equations are solved, the values of $D_\bot$ and $\xi$
should again be deduced from Eqns.
(\ref{eq:defM}--\ref{eq:difmacro}).

We start with showing that the solution of Eqn. (\ref{eq:legen0}) with
$\partial 
f /\partial x = 0$  again takes the form
\begin{equation}
g_0(p)=\Nu{p}-\Nd{p} \,.
\label{eq:zerosol}
\end{equation}
The difference from zero-temperature case (see
Eqn. (\ref{eq:mageigen})) is that the Fermi functions on the
r.\ h.\ s.\ of Eqn. (\ref{eq:zerosol}) are now given by
\begin{equation}
n^{\uparrow,\downarrow}_{\vec{p}}=
\frac{1}{\exp \left[ \frac{1}{T} \left( \frac{p^2}{2m} \mp
 \frac{H}{2} - \mu \right) \right] +1}
\label{eq:fermiT}
\end{equation}
rather than by simple step functions.

Since for the function $g_0(p)$ given by Eqn. (\ref{eq:zerosol})
the l.\ h.\ s.\ of Eqn. (\ref{eq:legen0}) vanishes, and so does
the operator $I_{sr}[g_0]$ on the r.\ h.\ s. (see
Eqn. (\ref{eq:spinrotcoll})) we have to prove only that the function
$g_0(p)$ solves the equation
\begin{equation}
I_{rel}[g(p)]=0\,.
\label{eq:rel0}
\end{equation}
Using Eqns. (\ref{eq:reltot}) and
(\ref{eq:aaproxT}),  we
rewrite (\ref{eq:rel0}) in terms of dimensionless variables,
\begin{equation}
\eta(p)= \frac{p^2}{2mT} - \frac{\mu}{T} \,,\,\,\,\,
h=\frac{ H}{2T}\,
\label{eq:dimensionless}
\end{equation}
as  
\begin{eqnarray}
g(\eta)(\eta^2+h^2+\pi^2)&-&\int_{-\infty}^{\infty}g(\sigma)(\eta+\sigma)
\times 
\label{eq:rel0explicit} \\
&\times& \left(\frac{1}{e^{\eta+h}+1} + \frac{1}{e^{\eta-h}+1} +
\frac{2}{e^{\eta+ \sigma}-1} \right) d\sigma = 0 \,,
\nonumber
\end{eqnarray}
where $g(\eta(p)) \equiv g(p)$.
It is convenient to transform Eqn. (\ref{eq:rel0explicit}) into a
differential equation  
\begin{eqnarray}
g''(k)& -& (h^2 + \pi^2)g(k) + 2 \pi^2 {\rm
sech}^2 \pi k \, g(k)= \nonumber \\
 &=& 2\pi \parallel g \parallel \frac{\sinh
\frac{h}{2}}{\cosh^2 \pi k} \,\left(\pi \sin kh\, \sinh \pi k -
 h \cos kh \cosh \pi k \right)
\label{eq:reldiff}
\end{eqnarray}
for the function
\[g(k) = \int_{-\infty}^{\infty} e^{ik\eta} g(\eta) \cosh \frac{\eta}{2}
\, d\eta\,,\]
with the normalization constant $\parallel g \parallel$  given by
\[\parallel g \parallel =\int_{-\infty}^{\infty}g(\eta)d\eta\,.
\]
Using a function $g_0(p)$ of the form (\ref{eq:zerosol}), we obtain
\begin{equation}
g_0(\eta)= \frac{1}{e^{\eta-h}+1} - \frac{1}{e^{\eta+h}+1}\,
\end{equation}
and the corresponding Fourier transform,
\begin{equation}
g_0(k)=\frac{2 \pi \sinh \frac{h}{2}\,\cos kh}{\cosh \pi k}\,.
\end{equation}
Finally, verifying that $g_0(k)$ is a solution
of Eqn. (\ref{eq:reldiff}) proves
that the function  (\ref{eq:zerosol}) satisfies Eqn. (\ref{eq:rel0}).

We can then re-write Eqn.
(\ref{eq:legen1}) in the steady state limit as
\begin{equation}
G_2 v_F\cdot (\Nu{p} - \Nd{p}) - i \Omega^{(1)}
f(p) = -I_{rel1}[f(p)]-iI_{sr1}[f(p)]\,.
\label{eq:difmicro}
\end{equation}
Here $v_F$ is the Fermi velocity and
$\Omega^{(1)}= 2a p_F H /\pi$ (cf. Eqn. (\ref{eq:omega1T0})).

With  the help of Eqns. (\ref{eq:reltot}) and (\ref{eq:aaproxT})
we obtain the explicit expression for
the integral of the relaxational term on the r.\ h.\ s. of
Eqn. (\ref{eq:difmicro}): 
\begin{equation}
\int_0^\infty I_{rel1}[W(p)]p^3 dp \approx
p_F^3\int_0^{\infty}
F(T,H,p)\,W(p) dp\,,
\label{eq:tauvargeneral} 
\end{equation}
\begin{equation}
F(T,H,p) =  \frac{8ma^2}{3\pi} \left[\left(\frac{p^2}{2m} -\mu
\right)^2 + \frac{1}{4}\, H^2 +
\pi^2 T^2 \right]\,\,.
\label{eq:FT} 
\end{equation}
Here the function $W(p)$ is assumed to be localized in the vicinity of
the Fermi surfaces, where $v_F \mid p - p_F \mid \stackrel{<}{\sim}
{\rm max} (T,H)$.

This yields the expected estimate $I_{rel1}[f(p)] \sim ma^2f(p)\, {\rm
max}(H^2,T^2)$ for the relaxational term on the r.\ h.\ s. of
Eqn. (\ref{eq:difmicro}).
In the case of spin-rotation term, 
the functions $n_{\vec{p}}^\uparrow - n_{\vec{p}}^\downarrow $
in Eqn. (\ref{eq:spinrotcoll}) carry a factor of $H$, and thus
$I_{sr1}[f(p)] \sim (p_F a)^2H f(p) \ll \Omega^{(1)}
f(p)$.

The ratio of the precession term $\Omega^{(1)} f(p)$ to the
relaxational term is always large for $H \stackrel{>}{\sim} T$, whereas
for $H \ll T$ it
is of the order of $\zeta$ (see Eqn. (\ref{eq:zeta})) and can vary
from $\zeta \gg 1$ (high-field region) to $\zeta \ll 1$ (low-field
region). On the other hand, the field and temperature dependence of
relaxational term itself is determined by the ratio $H/T$. 
Below we will use two different methods to solve
Eqn. (\ref{eq:difmicro}) in two regions, $\zeta \gg 1$ and $H/T \ll
T/\epsilon_F$, assuming in both cases that $H,T \ll \epsilon_F$.
Owing to the presence of a large factor $(ap_F)^{-1}$ in the definition 
of $\zeta$, the two regions overlap. Since the highly-polarized case
$H \gg T$ (with no requirement $H \ll \epsilon_F$) is described by the $T
\rightarrow 0$ limit of Sect. \ref{sec:zeroT}, we will thus cover the
entire degenerate region ($T \ll \epsilon_F$) for all values of $H$.

\underline{High-Field Behaviour, $\zeta \gg 1$:} 
In this case
the relaxational term in (\ref{eq:difmicro}) is small,
$I_{rel1}[f] \ll \Omega^{(1)}f$, and the $\tau$-approximation estimate
for the spin-rotation parameter (see Eqn. (\ref{eq:tauapprox})) yields
$\xi \gg 1$, so that 
in the spin-space the current $J_\bot$ is almost perpendicular to the
magnetization gradient.
In this limit the solution of Equation (\ref{eq:difmicro}), $f_D (p)$,
can be obtained iteratively in complete analogy with the
zero-temperature case of Sect. \ref{sec:zeroT}:
\begin{equation}
f_D (p)=-i\frac{v_F}{\Omega^{(1)}}G_2 \,
(\Nu{p}-\Nd{p}) - \frac{v_F}{(\Omega^{(1)})^2}G_2\,
I_{rel1}[\Nu{p}-\Nd{p}]-i\frac{v_F} {(\Omega^{(1)})^2}
G_2\, I_{sr1}[\Nu{p}-\Nd{p}]
\,\,.
\label{eq:highsol}
\end{equation}
Upon integration this leads to
\begin{equation}
J^-_\bot=-i\frac{v_F}{\Omega^{(1)}}G_2\,
J_\bot [\Nu{p}-\Nd{p}]\left\{ 1- \frac{1}{\Omega^{(1)}}
 \Omega^{(2)}[\Nu{p}-\Nd{p}] - 
 \frac{i}{\Omega^{(1)}}\,\frac{1}{ \tau_\bot[\Nu{p}-\Nd{p}]}\right\}
\,,
\label{eq:curhigh}
\end{equation}
where $J_\bot[W(p)]= \int W(p) p^3 dp / (12 \pi^2 m)$ (see
Eqn. (\ref{eq:defJ})), 
and the functionals $\Omega^{(2)}[W(p)]$ and $\tau_\bot[W(p)]$ are
defined by
\begin{equation}
\tau_\bot[W(p)]=\frac{\displaystyle \int_{0}^{\infty}
W(p) p^3 dp}  
{\displaystyle \int_{0}^{\infty}I_{rel1} [W(p)] p^3 dp}  
\,,\,\,\,\,\,\,\,\,
\Omega^{(2)}[W(p)]=-\frac{\displaystyle
\int_{0}^{\infty}I_{sr1} [W(p)] p^3 dp} 
{\displaystyle \int_{0}^{\infty} W(p) p^3 dp}\,.
\label{eq:deftau}
\end{equation}
The values of these functionals at
the exact leading-order solution, $\Nu{p}-\Nd{p}$,  represent 
respectively  the second order correction to the spin current
precession frequency and the ``diffusion time" $\tau_D$.
Calculating the value of $\tau_D$ with the help of
Eqns. (\ref{eq:tauvargeneral}--\ref{eq:FT}) and performing explicit
integration in 
$\Omega^{(2)}$, we arrive at \cite{misprint}: 

\begin{eqnarray}
D_\bot&=&\frac{v_F^2 \tau_D}{3}= \frac{3 \pi v_F^2}{8 m a^2 ( H^2 +
4 \pi^2 T^2)} 
\label{eq:diffhigh}\\
\xi&=& (\Omega^{(1)} + \Omega^{(2)}) \tau_D = \frac {9 H
v_F} {4 a (H^2 +
4 \pi^2 T^2)} [1 + \frac{4}{5 \pi^2}a p_F (1- 2 \ln 2)].
\label{eq:xihigh}
\end{eqnarray}

The first thing to stress is that in the entire high-field region
the value of $D_\bot$ in (\ref{eq:diffhigh}) differs 
from that of the longitudinal spin diffusion coefficient \cite{Miyake} ,
\begin{equation}
D_\parallel=\frac{v_F^2 \tau_\parallel}{3} \approx (v_F^2 / 8 \pi m
a^2 T^2 ) C(-1/3)\,, 
\label{eq:longitudinal}
\end{equation}
where $C(-1/3) \approx 0.843$ is the Brooker--Sykes
coefficient~\cite{Brooker}.
The two diffusion coefficients,  $D_\bot$
and $D_\parallel$, are plotted in Fig. \ref{fig:diffhigh}. 
Note that the spin
diffusion remains anisotropic even at $T \gg H$ (provided that $\zeta
\gg 1$), in which case we obtain from (\ref{eq:diffhigh}):
\begin{equation} 
D_\bot \approx \frac{3 v_F^2 }{ 32 \pi m a ^2 T^2} \approx 0.890
D_\parallel
\label{eq:approx089}
\end{equation}
in agreement with Ref. \cite{Pal}.
The difference between $D_\bot$ and $D_\parallel$ at $\zeta \gg
1$ reflects the fact that in this regime the spin rotation effects
dominate, as can be seen from the leading role of the
precession term $-i\Omega^{(1)}f(p)$ in Eqn. (\ref{eq:difmicro}).
Obviously, no such term enters the Boltzmann equation for the
longitudinal spin diffusion, and the longitudinal spin current is
by definition parallel to the gradient of $M_\parallel$. Thus in spite
of the identical temperature dependence of $D_\bot$ and $D_\parallel$,
the nature of the spin 
transport differs in the two cases.

The crossover between $T^{-2}$ and $H^{-2}$ behaviours of $D_\bot$ 
occurs at $T
\sim H$. Eqn. (\ref{eq:diffhigh}) shows that the
diffusion rate $\tau_D^{-1}$ obeys ``Matthiessen's rule'',
\begin{equation}
\frac{1}{\tau_D}=\frac{1}{\tau_H}+\frac{1}{\tau_T}\,,
\label{eq:matthiessen}
\end{equation}
where the quantities
\begin{equation}
\tau_H=\frac{9 \pi}{8 ma^2 H^2}\,,\,\,\,\,\,\,
\tau_T=\frac{9}{32 \pi m a^2 T^2} \approx 0.890 \tau_\parallel
\end{equation}
can be thought of as the relaxation times for field- and
temperature-dominated collision processes. 

We note that it has been commonly accepted
\cite{Mullin89,Mullin92,Karen3} that the crossover of
$D_\bot$ at $H \sim T$ corresponds to transition from anisotropic to
isotropic behaviour of spin diffusion. We will see that in reality this
transition occurs at much lower fields, $H \sim (ap_F)T^2/\epsilon_F$.

The dotted line in Fig. \ref{fig:diffhigh} corresponds to the quadratic
fit \cite{Candela,Karen3} 
\begin{equation}
\tilde{D}_\bot =\frac{\pi v_F^2}{8 m a^2} \left\{ \frac{1}{3}H^2 +
C(-1/3)\, T^2 \right\}^{-1}
\label{eq:difffit}
\end{equation}
which is obtained by substitution $\tau_T \rightarrow \tau_\parallel$ in
Eqn. (\ref{eq:matthiessen}) and is widely used to describe the
temperature dependence of $D_\bot$. One can see that while this fit
provides correct order-of-magnitude estimate, it deviates from the
exact result (\ref{eq:diffhigh}) at $T \stackrel{>}{\sim} H$.
 
\underline{Crossover and Low-Field Behaviour, $H/T \ll T/\epsilon_F$:}
In this region, the spin-rotation term on the r.\ h.\ s.\ of the
steady-state equation  (\ref{eq:difmicro}) is small in comparison to
both the relaxational and precession terms,
\begin{equation}
 I_{sr1}[f(p)] \ll I_{rel1}[f(p)] \,,\,\,\,\,\,\,
 I_{sr1}[f(p)] \ll \Omega^{(1)}f(p)\,,
\end{equation}
and thus, it may be omitted in the leading-order calculation.
Furthermore, one may also set $H=0$ in evaluating
the collision term $I_{rel1}$.

In terms of the reduced variables (\ref{eq:dimensionless}),
Eqn. (\ref{eq:difmicro}) then becomes
\begin{eqnarray}
-\frac{v_F}{T^2} G_2&&\left(\frac{1}{e^{\eta-h}+1}
-\frac{1}{e^{\eta+h}+1} \right) = \frac{2a^2m}{\pi}\,f(\eta) \left(
\eta^2 + \pi^2 - 
\frac{i \pi}{2 a^2 m T^2} \Omega^{(1)} \right) - \nonumber \\
&&-\frac{4a^2m}{3\pi}
\int_{-\infty}^{\infty}(\eta+\sigma)f(\sigma)\left(\frac{1}
{e^\sigma+1}+
\frac{1}{e^{\eta+\sigma}-1} \right)d\sigma\,,
\label{eq:diffintermed}
\end{eqnarray}
where 
$f(\eta(p)) \equiv f(p)$ and $h \ll 1$.
Since the l.\ h.\ s. of this
equation is even in $\eta$ and the integral term on the r.\ h.\ s.\
has the parity of the function $f(\eta)$, the inhomogeneous equation
(\ref{eq:diffintermed}) should have an even solution
$f(\eta)=f(-\eta)$.
In finding this solution, we will make use of the methods of
Ref. \cite{Brooker} and begin by
transforming (\ref{eq:diffintermed}) into the differential
equation

\begin{equation}
F''(k) - \pi^2 \gamma^2 F(k) - \frac{2}{3} {\rm sech}^2 \pi k \,F(k) =
\frac{\pi^2 h v_F}{2 a^2 m T^2} G_2\,
\frac{\cos k h}{\cosh \pi k}
\label{eq:intermed2}
\end{equation}
for the function
\[F(k) = \int_{-\infty}^{\infty}e^{i k \eta} f(\eta) \cosh
\frac{\eta}{2} \, d \eta \,\,,\,\,\,\,\,\,\,\,F(k)=F(-k).\]
Here
\[\gamma^2=1- \frac{i \Omega^{(1)}}{2 \pi a^2 m T^2} \approx
1 - 2 i \zeta /\pi^2
\,.\]
The solution of Eqn. (\ref{eq:intermed2}) can be written as
an expansion
\begin{equation}
F(k) = \sum_{n=0}^\infty F_n \phi_n(k)
\label{eq:expansionF}
\end{equation}
in the basis provided by the functions 
\begin{equation}
\phi_n(k)=(1-\chi^2)^{\gamma/2}C^{\gamma +
\frac{1}{2}}_n(\chi)\,,\,\,\,
\chi= \tanh \pi k\,,
\label{eq:diffhomeigen}
\end{equation}
where $C^{\gamma +\frac{1}{2}}_n(\chi)$ are Gegenbauer polynomials
with a complex index $\gamma + 1/2$. The functions $\phi_n(k)$
satisfy the homogeneous equations
\begin{equation}
\phi''(k) - \pi^2 \gamma^2 \phi(k) - (\gamma+n)(\gamma+n + 1){\rm
sech}^2 \pi k \,\phi(k) =0
\label{eq:diffhomogen}
\end{equation}
and are orthogonal to each other,
\begin{equation}
\int_{-\infty}^{\infty}\phi_n(k) \phi_m(k) {\rm sech^2} \pi k \,dk =
\frac{2^{-2\gamma}\Gamma(2\gamma+n+1)}{n{\rm !}\,(n+\gamma+\frac{1}{2})
\left[\Gamma(\gamma+\frac{1}{2})\right]^2}\delta_{mn}\,.
\end{equation}
Since $\phi_n(-k) = (-1)^n \phi_n(k)$, only the
functions $\phi_n(k)$ with even $n$ contribute to the expansion 
(\ref{eq:expansionF}) of an even function $F(k)$.
Substituting the expansion
\begin{equation}
\frac{\pi}{3} \,\frac{\cos kh} {\cosh \pi k} \approx \frac{\pi}{3}
\,\frac{1}{\cosh \pi k}={\rm sech^2} \pi k \, 
\sum_{n=0}^\infty g_n \phi_n(k)\,,
\label{eq:diffexpansion}
\end{equation}
where the summation again has to be performed over even $n$ only, and 
\begin{equation}
g_n=\frac{\pi^{5/2}(n+\gamma + \frac{1}{2})\Gamma(\gamma+\frac{1}{2})}
{3 \cos \frac{\pi \gamma}{2}\, \Gamma(\frac{1}{2}-\frac{n}{2}) 
\Gamma(1+\gamma+\frac{n}{2})\Gamma(1+\frac{\gamma}{2}+\frac{n}{2})
\Gamma(\frac{1}{2}-\frac{\gamma}{2}-\frac{n}{2})}
\end{equation}
into the r.\ h.\ s.\ of Eqn. (\ref{eq:intermed2})  we finally obtain
\begin{equation}
F_n = -\frac{3 v_F h}{2 \pi a^2 m T^2}G_2\,
\frac{g_n}{\frac{2}{3}+ (\gamma+n)(\gamma+n+1)}\,
\label{eq:inhomeigen}
\end{equation}
for even $n$.
As was already mentioned above,  $F_n=g_n=0$ for odd $n$.

The transverse spin current (\ref{eq:defJ}) can now be evaluated as 
\begin{equation}
J_\bot^-=\frac{T p_F^3}{12\pi^2}\int_{-\infty}^\infty \frac{F(k)}{\cosh
\pi k} dk = \frac{T p_F^3}{12\pi^2} \sum_{n=0}^\infty F_n a_n\,;
\label{eq:curintermed}
\end{equation}
\begin{eqnarray}
a_n &\equiv& \int_{-\infty}^\infty\frac{\phi_n(k)}{\cosh \pi k} dk = \\
&=&\frac{\pi \Gamma(\gamma+1) \Gamma(2\gamma+n+1)}{n{\rm !} \cos\frac{\pi
\gamma}{2}\, \Gamma(2\gamma+1) \Gamma(\frac{1}{2}-\frac{n}{2})
\Gamma(\gamma+\frac{n}{2}+1)\Gamma(\frac{\gamma}{2}+\frac{n}{2}+1)
\Gamma(\frac{1}{2}-\frac{n}{2}-\frac{\gamma}{2})}\,
\nonumber
\end{eqnarray}
for even $n$,
and $a_n=0$ for odd $n$.
Finally, Eqns. (\ref{eq:ansatzg} -- \ref{eq:difmacro}) and
(\ref{eq:zerosol})  yield the expression
\begin{eqnarray}
\frac{D_\bot}{1-i\xi} &=& -\frac{v_F T}{3  H G_2} 
\sum_{n=0}^\infty F_n a_n =
\nonumber \\
&=&\frac{1}{4 \pi}\frac{v_F^2}{m a^2 T^2} \sum_{n=0}^\infty
\frac{g_n a_n}{\frac{2}{3} +(\gamma+n)(\gamma+n+1)}\,,
\label{eq:intermedanswer}
\end{eqnarray}
from which one can extract  $D_\bot$ and spin-rotation
parameter $\xi$ (see Fig. \ref{fig:diffcross}).
In the high-field limit $\zeta \gg 1$,
Eqn. (\ref{eq:intermedanswer}) gives 
\begin{equation}
D_\bot \approx 0.890 D_\parallel\,,\,\,\,\,\,\,\,
\xi= \frac{9}{8 \pi^2} \zeta \gg 1
\label{eq:abovecross}
\end{equation}
in agreement with Eqns. (\ref{eq:approx089}) and (\ref{eq:xihigh}),
while in the low-field case of $\zeta \ll 1$ we obtain the isotropic result,
\begin{equation}
D_\bot =D_\parallel\,,\,\,\,\,\,\,\,\,\,
\xi \approx 0.139 \zeta \ll 1\,\,.
\label{eq:belowcross}
\end{equation}
Note that while the value of 
$D_\bot$ in (\ref{eq:belowcross}) can be obtained from the
$\tau$-approximation formula (\ref{eq:tauapprox}) with $\tau_\bot =
\tau_\parallel$ (see Eqn. (\ref{eq:longitudinal})), spin-rotation
parameter $\xi$ is {\em not} given by $\Omega^{(1)} \tau_\parallel$.

The crossover between Eqns. (\ref{eq:abovecross}) and
(\ref{eq:belowcross}) at $\zeta \sim 1$ cannot be properly accounted
for within the 
relaxation-time approximation (Eqns. (\ref{eq:tauapprox})) because
the relaxation operator $I_{rel1}$ on the r.\ h.\ s. of
Eqn. (\ref{eq:difmicro}) (see also Eqns.
(\ref{eq:tauvargeneral}--\ref{eq:FT})) does not depend on the magnetic
field for $H \ll T$. This is in contrast with the interpretation of the 
low field
crossover as a result of the competition between two different terms in the
relaxational part of the collision operator \cite{Sergey}. 
As illustrated in Fig. \ref{fig:changeform}, what changes at $\zeta \sim
1$ is {\em the form of the solution}, $f(p)$, to the steady-state
equation (\ref{eq:difmicro}). It is this change in $f(p)$ which is not
accounted for within the relaxation time approximation, hence the
absence of the second crossover in that approach \cite{Karen3}.

Although quantitatively our considerations apply only to the
degenerate case ($T \ll \epsilon_F$), there is no reason to doubt that
the crossover at $\xi \sim 1$ should also be present 
in the case of $H \stackrel {>}{\sim} (a p_F) \epsilon_F$ and 
$T \stackrel {>}{\sim} \epsilon_F$, in the presence of higher partial
waves, or in
a system with stronger interactions.
Since the separation of scales of the two parameters is
provided by the factor $(\epsilon_F /T)/ (a p_F) \gg 1$ in Equation
(\ref{eq:zeta}), in both of these cases the intermediate region
between the two crossovers  should shrink. In the next section, we
will discuss the experimental results that support this point of view.

\section{Discussion of Experimental Results}
\label{sec:discu}

Some comments are in order concerning the possible 
relevance of our findings to experiment.
Even though, strictly speaking, our analysis does not apply 
to the strongly interacting
case,
it is worth noting that the available experimental data
in weakly polarized $^3{\rm He}$ ~\cite{Candela}
deviate systematically from the simple theoretical fit
which uses a single adjustable parameter ($T_a$ in \cite{Candela})
to cover the entire temperature range including both $\zeta >1$ and
$\zeta < 1$. 
Much better agreement is obtained by restricting the fit to
the $\zeta \stackrel{>}{\sim} 1$ region with an overall prefactor smaller than
the one implied by fitting to the value of $D_\parallel$ in the
low field, high temperature regime (in re-analyzing the data  we use
the value $A \approx 7.7 \times 10^5 
\rm{sec}/\rm{cm}^2 \rm{K}^2$ 
instead of $5.8 \times 10^5 \rm{sec}/\rm{cm}^2 \rm{K}^2$ for the
inverse overall 
prefactor in Eqn. (2) of Ref. \cite{Candela}; we also take the value 
12.5 mK instead of 16.4 mK for the ``anisotropy temperature'' $ T_a$). This
is consistent with 
our picture, with $D_\bot < D_\parallel$ for 
$\zeta \stackrel{>}{\sim} 1$ (i.\ e. $\xi \stackrel
{>}{\sim}1$)  and $T \gg H$. In addition, 
although it appears that the region between the 
two crossovers cannot be clearly identified --  most likely due to large
Fermi liquid renormalization effects -- the isotropic limit is indeed reached
in the regime $\xi < 1$ \cite{Candela}. We also note the systematic
deviations from the quadratic fit of Ref. \cite{Karen3} visible in
the measurements of $D_\bot$ in a $6.4 \%$ $^3{\rm He}$ --$^4{\rm He}$
mixture \cite{Owers95} in the intermediate region between $T \sim T_a$
and $\mid\xi\mid \sim 
1$. However, the interactions in this system are still too strong for
our theory to be applicable. The fact that in both cases of
Refs. \cite{Candela} and \cite{Owers95} the intermediate region
appears to be rather narrow is in line with our expectations (see
Sect. \ref{sec:smallpol}).
In principle,
our calculations should be more relevant to the measurements in
dilute $^3{\rm He}$ --$^4{\rm He}$ mixtures.
Although in the available data 
(for $.18 \% \,\,^3{\rm He}$)
the crossover to the
isotropic limit occurs for $\xi \sim 1$ with $H \ll T$, the temperature is
not sufficiently far below $\epsilon _F$ and, moreover, the polarization is
somewhat high, $\sim 25 \%$.
Nevertheless, 
for $\xi > 1$ $D_\bot \propto D_\parallel$ with
the ratio $D_\bot / D_\parallel$ slightly less than 
unity \cite{Candela91,Nunes}. 
Also, the measured $T$ dependence of the ``spin-rotation"
parameter, $\xi$, near the crossover to the isotropic (``low-field")
limit 
is qualitatively consistent with our results in 
both the data of
Reference \cite{Candela91} and those obtained in the degenerate
regime of more concentrated solutions ($2.6 \%\,\, ^3{\rm
He}$) \cite{Owers} with lower 
polarization ($\sim 2 \%$). In both situations, the crossover to the
isotropic regime can be clearly distinguished.
However, in the former case the
polarization was again rather high ($\sim 25 \%$) and the
``low-field'' crossover 
itself occurs beyond the degenerate limit.
Also, there is a large discrepancy in the magnitude of 
the shift of $\xi T^2$ in the latter case (see Fig. \ref{fig:spinrot})
which can be attributed to Fermi liquid renormalizations anticipated
in high concentration solutions. 
The deviations of $\xi$ at $\xi \sim 1$ from the predictions of
Ref. \cite{Mullin92} are also seen in the data for the dilute ($.05 \%$ and $.1
\%$ $\,^3{\rm He}$ ) mixtures of Ref. \cite{Owers95}, in which case
due to the high polarization 
$\xi \sim 1$ corresponds to $T \sim
\epsilon_F$. Again, our results do not apply quantitatively to this
situation. In addition, the data points are somewhat too scattered
for a clear identification of the crossover.
To sharpen the identification of two crossovers 
the data of reference \cite{Owers} should be extended to lower temperatures
(to study the $H/T \sim 1$ behaviour). Quantitative comparisons could be made
only in the more dilute case of references \cite{Candela91,Owers95} where lower
field and lower temperature experiments should be performed.

\acknowledgements

We are grateful to V. A. Brazhnikov, A. E. Meyerovich, and K. A. Musaelian
for helpful and enjoyable discussions and to D. Candela for providing
us with the data of References \cite{Candela} and \cite{Candela91} 
in a convenient form. AER would like to thank the Alexander von Humboldt
Foundation for generous support in the form of a Senior 
Research Award, and Peter
W\"{o}lfle, Gerd Sch\"{o}n and Albert Schmid
for their kind hospitality in the condensed matter theory group in Karlsruhe,
where the final version of this paper was completed.  
This work was also supported in part by ONR Grant \# N00014-92-J-1378.

\appendix{On the Dynamics of Quasiparticles in a Spin-Polarized Fermi
Gas} 
\label{app:self}

In this Appendix, we derive 
the imaginary parts of quasiparticle
self-energies at $T=0$ which are used as input into the kinetic
equations. 

We start with a spin-polarized gas of neutral fermions interacting via
a repulsive potential $U(\mid \vec{r}- \vec{r}^{\,\prime} \mid)$. In the
dilute limit, $N^{1/3}r_0 \ll 1$, where $r_0$ is the range of the
potential, the leading contribution comes from the $s$-wave scattering
between spin-up and spin-down particles. Then the natural small
parameter is provided by $N^{1/3}a$, where $a$ is the $s$-wave
scattering length. We will keep terms up to the second order in
$N^{1/3}a$; within this accuracy, the scattering is fully described by
a two-body $T$-matrix which obeys the Bethe-Salpeter equation: 
\begin{equation}
T(\vec{p}_1,\vec{p}_2;(s_0,\vec{s})) = - U(\vec{p}_1-\vec{p_2})+ \int
\frac {d^3k}{(2\pi)^3}
\frac{T(\vec{p}_1,\vec{k};(s_0,\vec{s}))U(\vec{k}-\vec{p} 
_2) \Nud}
{2s_0 -s^{\,2}/m+ 2
\mu - k^2/m+i {\cal N}^{\uparrow \downarrow}(\vec{k})\cdot 0}\,,
\label{eq:Dyson}
\end{equation}
where 
\begin{equation}
{\cal N}^{\uparrow
\downarrow}(\vec{k})=1-n^\uparrow_{\vec{s}+\vec{k} }-n^\downarrow
_{\vec{s} - \vec{k}}\,\,\,\,,
\label{eq:Nud}
\end{equation}
$n^\uparrow_{\vec{p}}$ and $n^\downarrow_{\vec{p}}$
are the Fermi distribution functions for spin-up and spin-down particles
respectively, and $(s_0,\vec{s}) $ is the 4-momentum
of the center of mass. The momenta of the incoming (outgoing) particles 
are $\vec{s}\pm\vec{p}_1$, 
($\vec{s}\pm\vec{p}_2$). Equation (\ref{eq:Dyson}) is solved by
\begin{eqnarray}
T(\vec{p}_1,\vec{p}_2;(s_0,\vec{s}))& =&
-\frac{4\pi a}{m} - \left(\frac{4\pi a}{m}\right)^2\int\left[ 
 \frac{{\cal N}^{\uparrow \downarrow}(\vec{k})}
{2s_0 -s^{\,2}/m+ 2
\mu - k^2/m+i {\cal N}^{\uparrow \downarrow}(\vec{k})\cdot 0}
-\right.   \nonumber \\
& &\left.
-{\cal P} \frac{m}{p^2_2-k^2}
\right] \frac {d^3k}{(2\pi)^3}\,\,\,\, ,
\label{eq:Tequil}
\end{eqnarray}
which is a straightforward generalization of a well-known result
\cite{Galitsky} for the case of a spin-polarized system. 

To obtain the self energies of spin-up ($\Suu{}(\omega,\vec{p})$) and
spin-down ($\Sdd{}(\omega,\vec{p})$) particles, one has
to perform the integration of the $T$-matrix multiplied by
the single-particle Green's 
function, as shown in Fig. \ref{fig:self}. 
We will first consider the
on-shell self-energies $\Sigma_{\uparrow
\uparrow}(\omega=\epsilon_\uparrow(\vec{p}),{\vec{p}})$ and
$\Sigma_{\downarrow
\downarrow}(\omega=\epsilon_\downarrow(\vec{p}),{\vec{p}})$, where
\begin{equation}
\epsilon_\uparrow(\vec{p})= \frac{p^2}{2m} - \frac{1}{2} H -
\mu,
\,\,\,\,\,\,\,
\epsilon_\downarrow(\vec{p})= \frac{p^2}{2m} + \frac{1}{2} H -
\mu\,\,\,\,
\label{eq:energies}
\end{equation}
are the single-particle energies.
To second order in $aN^{1/3}$ we obtain
\begin{eqnarray}
&& \Sigma_{\uparrow \uparrow}({\vec{p}}) \equiv  \Sigma_{\uparrow
\uparrow}(\epsilon_\uparrow(\vec{p}),{\vec{p}})  =  \frac{4\pi a}{m}
N_\downarrow + \label{eq:sigup} \\
&&+\frac {(4\pi a)^2}{2m} \int \left[ \frac{{\cal N}^{\uparrow
\downarrow}(\vec{k})({\cal N}^{\uparrow \downarrow}(\vec{k})+
2n^\downarrow_{{\vec{p}}^{\,\prime}}- 1) }
{\frac{1}{4}(\vec{p}-\vec{p}^{\,\prime})^2-k^2
+ im{\cal N}^{\uparrow \downarrow}(\vec{k}) \cdot 0} - {\cal P} 
\frac{2n^\downarrow_{{\vec{p}}^{\,\prime}}}
{\frac{1}{4}(\vec{p}-\vec{p}^{\,\prime})^2-k^2}
\right] 
\frac{d^3k\,d^3p^\prime}{(2\pi)^6}\,\,\,, \nonumber\\
& & \Sigma_{\downarrow \downarrow}({\vec{p}}) \equiv  \Sigma_{\downarrow
\downarrow}(\epsilon_\downarrow(\vec{p}),{\vec{p}})  =  \frac{4\pi a}{m}
N_\uparrow + \label{eq:sigdown} \\
&& +\frac {(4\pi a)^2}{2m} \int \left[ \frac{{\cal N}^{\uparrow
\downarrow}(\vec{k})({\cal N}^{\uparrow \downarrow}(\vec{k})+
2n^\uparrow_{{\vec{p}}^{\,\prime}}- 1) }
{\frac{1}{4}(\vec{p}-\vec{p}^{\,\prime})^2-k^2
+ im{\cal N}^{\uparrow \downarrow}(\vec{k}) \cdot 0} - {\cal P} 
\frac{2n^\uparrow_{{\vec{p}}^{\,\prime}}}
{\frac{1}{4}(\vec{p}-\vec{p}^{\,\prime})^2-k^2}
\right] 
\frac{d^3k\,d^3p^\prime}{(2\pi)^6}\,\,\,. \nonumber
\end{eqnarray}
Here the center-of-mass momentum $\vec{s}$ that enters the definition of \Nud
(see Eqn. (\ref{eq:Nud})) is equal to
$(\vec{p}+\vec{p}^{\,\prime})/2$.

Separating real and imaginary parts
in Eqn. (\ref{eq:sigup}) yields
\begin{eqnarray}
\R\Sigma_{\uparrow \uparrow}(\vec{p}) = \frac{4\pi
a}{m}N_\downarrow + \frac{(8\pi a)^2}{m}\int{\cal P}\left[
\frac{n^\uparrow_{\vec{s}+\vec{k}} n^\downarrow_{\vec{s}-\vec{k}} -
 n^\downarrow_{\vec{p}^{\,\prime}}n^\uparrow_{\vec{s}+\vec{k}}-
 n^\downarrow_{\vec{p}^{\,\prime}}n^\downarrow_{\vec{s}-\vec{k}}}
{(\vec{p}-{\vec{p}}^{\,\prime})^2-4k^2} 
\right]\frac{d^3k\,d^3p^\prime}{(2\pi)^6}
\label{eq:resigup}
\end{eqnarray}
and
\begin{eqnarray}
\I\Sigma_{\uparrow \uparrow}(\vec{p})& = &
\pi\frac{(4\pi a)^2}{m}\int\left[
n^\uparrow_{\vec{s}+\vec{k}} n^\downarrow_{\vec{s}-\vec{k}} -
 n^\downarrow_{\vec{p}^{\,\prime}}
\left( \Nud +
2n^\uparrow_{\vec{s}+\vec{k}} n^\downarrow_{\vec{s}-\vec{k}} \right)
\right] \times \nonumber \\
&&\times\delta \left(k^2-\frac{(\vec{p}-\vec{p}^{\,\prime})^2}{4}
 \right) 
\frac{d^3k\,d^3p^\prime}{(2\pi)^6}\,\,\,;
\label{eq:imsigup}
\end{eqnarray}
similar equations with $\uparrow \leftrightarrow \downarrow$ hold for 
$\R\Sdd{}$ and $\I\Sdd{}$.
In the
non-polarized case, when $n^\uparrow_{\vec{p}}=n^\downarrow_{\vec{p}}\,\,$, the
expressions (\ref{eq:sigup}--\ref{eq:imsigup}) reduce
to the familiar formulae for the self energy \cite{Galitsky,Volume9}.

At $T=0$, it is possible to 
perform the integration in Eqn. (\ref{eq:imsigup}) (and in the
similar expression for $\I\Sigma_{\downarrow \downarrow}$)
explicitly. After a cumbersome calculation we obtain
the following results for the imaginary parts of  self
energies:

\noindent At $q>\Pu$,
\begin{eqnarray}
\I\Sigma_{\uparrow \uparrow}(\vec{q})& = &-\frac{a^2}{15 \pi q m}\left[
5q^2\Pd^3-2\Pd^5-5\Pu^2\Pd^3+ \right. \nonumber \\
&&+\left. 2\theta(\Pd^2+\Pu^2-q^2)\cdot (\Pd^2+\Pu^2-q^2)^{5/2}
\right] \,\,\,\,, \label{eq:sigupo} \\
\I\Sigma_{\downarrow \downarrow}(\vec{q})& = &-\frac{a^2}{15 \pi q m}\left[
5q^2\Pu^3-3\Pu^5+\Pd^5-5\Pu^2\Pd^3+ \right. \nonumber \\
&&+\left. 2\theta(\Pd^2+\Pu^2-q^2)\cdot (\Pd^2+\Pu^2-q^2)^{5/2}
\right] \,\,\,\,. \label{eq:sigdowno}
\end{eqnarray}
In the intermediate region $\Pd<q<\Pu\,\,\,$,
\begin{eqnarray}
\I\Sigma_{\uparrow \uparrow}(\vec{q})& = &\frac{a^2}{15 \pi q m}\left[
-5q^2\Pd^3-2\Pd^5+5\Pu^2\Pd^3+ \right. \nonumber \\
&&+\left. 2\theta(q^2-\Pu^2+\Pd^2)\cdot (q^2-\Pu^2+\Pd^2)^{5/2}
\right] \,\,\,\,, \label{eq:sigupm} \\
\I\Sigma_{\downarrow \downarrow}(\vec{q})& = &-\frac{a^2}{15 \pi q m}\left[
2q^5-5q^2\Pd^3+3\Pd^5\right]\,\,\,.\label{eq:sigdownm}
\end{eqnarray}
And, finally, in the innermost region $q<\Pd$,
\begin{eqnarray}
\I\Sigma_{\uparrow \uparrow}(\vec{q})& = &\frac{a^2}{60 \pi q m}\left[
7q^5-20q^3\Pd^2-10q^3\Pu^2-15q\Pd^4+30q\Pd^2\Pu^2+ \right. \nonumber \\
&&\left. +8\theta(q^2-\Pu^2+\Pd^2)\cdot (q^2-\Pu^2+\Pd^2)^{5/2}
\right] \,\,\,\,, \label{eq:sigupi} \\
\I\Sigma_{\downarrow \downarrow}(\vec{q})& = &\frac{a^2}{4\pi  m}
(\Pd^2-q^2)^2\,\,\,\,. \label{eq:sigdowni}
\end{eqnarray}
Equations (\ref{eq:sigupo}) and
(\ref{eq:sigdowni}) coincide with those reported in
Ref. \cite{KarenPR}.
The expressions (\ref{eq:sigdowno}--\ref{eq:sigupi}) are presented
here for the first time.

The singularity at $q^2=\Pu^2+\Pd^2$, which is present at Eqns.
(\ref{eq:sigupo}--\ref{eq:sigdowno}) corresponds to a process in which
an incoming particle excites a particle of opposite spin from within
the Fermi sphere. At $q>(\Pu^2+\Pd^2)^{1/2}$, the incoming particle has
enough energy to excite {\em any\/} particle of opposite
spin. Geometrically, this singularity corresponds  to the exterior
tangency of the shifted
Fermi spheres, and it is already present
in the unpolarized case \cite{Galitsky}.

In the spin-polarized case there is one more singularity at
$q^2=\Pu^2-\Pd^2$ (see Eqns. (\ref{eq:sigupm} ) and
(\ref{eq:sigupi})), associated with the interior tangency between the
two Fermi spheres. In this case, the corresponding virtual process is a
recombination of the incoming spin-up hole, followed by an excitation
of spin-down particle; at $q<(\Pu^2-\Pd^2)^{1/2}$, the hole has enough
energy to excite {\em any\/} of the spin-down particles from within
the spin-down Fermi sphere.

The functions $\I\Sigma_{\uparrow \uparrow}(\vec{q})$ and 
$\I\Sigma_{\downarrow \downarrow}(\vec{q})$ change their signs at
$q=\Pu$ and $q=\Pd$ respectively. The new feature of the above
results, relevant for the transverse spin dynamics, is that, due to
the availability of the scattering phase space between the two Fermi
surfaces, the lifetimes of spin-down
particles and spin-up holes in the intermediate region $p_\downarrow <
q < p_\uparrow$ are finite.

The transverse spin processes are described by the 
Wigner transform of the transverse component of the density matrix, 
\Nt{q}, which obeys the Boltzmann equation of a peculiar type (see
Sect. \ref{sec:zeroT}--\ref{sec:smallpol}). In the appropriate
linearized approximation sufficient for the calculation of transport
coefficients the transverse components
decouple from the longitudinal ones,
enabling one to discuss the transverse and
longitudinal spin dynamics separately \cite{Helium3}. 
By rotational symmetry, the transverse deviations from the equilibrium
(uniform) polarization must be small at any point in $\vec{p}$ space,
i.\ e.
\begin{equation}
\Nt{q} \ll \Nu{q}-\Nd{q}.
\label{eq:domain1}
\end{equation}
This also respects the fact that the eigenvalues of the density matrix
are between zero and unity.

Thus, in a non-equilibrium state the
value of \Nt{q} has to differ from zero over a finite interval of
values of $q$ within the intermediate region $p_\downarrow< q <
p_\uparrow$. The fact that the quasiparticle relaxation rates
(\ref{eq:sigupm}--\ref{eq:sigdownm}) also differ from zero within this
region results 
in the finite relaxation time $\tau_\bot$ of the transverse spin
current even at 
$T \rightarrow 0$ (see Sect. \ref{sec:zeroT}).

We note that from the viewpoint of longitudinal spin dynamics the
system behaves as a conventional Fermi liquid: the non-equilibrium terms in
the diagonal components of density matrix are confined to the
respective Fermi spheres, where the imaginary parts of the
quasiparticle spectrum vanish. As expected \cite{Sartor}, the
imaginary part of the off-shell self energy at the
Fermi surface behaves as
\begin{equation}
\I\Sigma_{\uparrow \uparrow}(\epsilon_\uparrow(\Pu)+\omega, \Pu)
\approx - {\rm sign}\omega\,\cdot\,\frac{a^2}{\pi m}\frac{\Pd}{\Pu}
\omega^2\,\,.
\end{equation}

\appendix{Derivation of Boltzmann Equations}
\label{app:kinetics}

In this Appendix, we derive the kinetic equation for transverse spin processes
using the Keldysh approach \cite{Volume10}. 
We will consider 
time dependent Green's functions of the form:
\begin{eqnarray}
iG^{--}_{1\alpha 2\beta}&= &\langle T \hat{\Psi}_{1\alpha} 
{{\hat{\Psi}^{\dag}}_{2\beta}}\rangle\,,\,\,\, 
iG^{++}_{1\alpha 2\beta}= \langle \tilde{T} \hat{\Psi}_{1\alpha}
{\hat{\Psi}^{\dag}}_{2\beta}\rangle\,, \nonumber \\
iG^{+-}_{1\alpha 2\beta}&= &\langle  \hat{\Psi}_{1\alpha} 
{\hat{\Psi}^{\dag}}_{2\beta}\rangle\,,\,\,\, 
iG^{-+}_{1\alpha 2\beta}= -\langle {\hat{\Psi}^{\dag}}_{2\beta}
 \hat{\Psi}_{1\alpha} \rangle\,,\label{eq:Green}
\end{eqnarray}
which satisfy the identity
\begin{equation}
iG^{--}_{1\alpha 2\beta}+iG^{++}_{1\alpha 2\beta}- 
iG^{+-}_{1\alpha 2\beta}-iG^{-+}_{1\alpha 2\beta}= 0\,\,.
\end{equation}
Here, the indices $1, 2$ stand for the two space-time arguments $X_i=
(t_i\,,\vec{r}_i\,)$,
$T$ and $\tilde{T}$ are the operators of chronological and
anti-chronological ordering respectively, and the average
$\langle \cdots \rangle$ is taken over the given non-equilibrium
state.

In the absence of the interaction, the functions (\ref{eq:Green})
obey the following equations of motion:
\begin{eqnarray}
\hat{G}^{-1}_{01\alpha\gamma}G^{(0)\,ab}_{1\gamma2\beta} &=&
\tau^{ab}_z\delta_{\alpha\beta}\delta(X_1-X_2)\,,\label{eq:dif1}\\
\left(\hat{G}^{-1}_{02\beta\gamma}\right)^\ast 
G^{(0)\,ab}_{1\alpha2\gamma} &=&
\tau^{ab}_z\delta_{\alpha\beta}\delta(X_1-X_2)\,,\label{eq:dif2}
\end{eqnarray}
where the indices $a,b$ accept values $+$ or $-$ and $\tau^{ab}_z$ is
a Pauli matrix: $\tau^{--}_z=-\tau^{++}_z=1$,
$\tau^{-+}_z=\tau^{+-}_z=0$. Here and below we presume that the sum
should be taken over the repeated Latin or Greek (i.\ e.\ spin)
indices. The operator $\hat{G}^{-1}_{01\alpha\beta}$ is defined as 
\begin{equation}
\hat{G}^{-1}_{01\alpha\beta}=(i\frac{\partial}{\partial t_1} +
\frac{\Delta_1}{2m} + \mu)\delta_{\alpha \beta} + \frac{1}{2} H
\sigma^z_{\alpha \beta} \,\,.\label{eq:dif3}
\end{equation}

The Green's functions of the interacting system satisfy
the Dyson equation
\begin{equation}
G^{ab}_{1\alpha2\beta}=G^{(0)ab}_{1\alpha2\beta}+\int
G^{(0)ac}_{1\alpha3\gamma}\Sigma^{cd}_{3\gamma4\lambda}
G^{db}_{4\lambda2\beta}\,d^4X_3\,d^4X_4\,\,,
\label{eq:sigdef1}
\end{equation}
or, equivalently,
\begin{equation}
G^{ab}_{1\alpha2\beta}=G^{(0)ab}_{1\alpha2\beta}+\int
G^{ac}_{1\alpha3\gamma}\Sigma^{cd}_{3\gamma4\lambda}
G^{(0)db}_{4\lambda2\beta}\,d^4X_3\,d^4X_4\,\,
\label{eq:sigdef2}
\end{equation}
where $\Sigma^{ab}_{1\alpha 2 \beta}$ are the proper self-energies.

Our aim is to examine the equation of motion for 
$G^{-+}_{1\uparrow2\downarrow}$ which is the off-diagonal component of
the density matrix in spin-space which is ultimately related to the
transverse magnetization and associated spin current (see
Eqns. (\ref{eq:defM} -- \ref{eq:defJ})).
To derive this equation we act on Eqns.
(\ref{eq:sigdef1}) and (\ref{eq:sigdef2}) with the operators
$\hat{G}^{-1}_{01\uparrow\uparrow}\,\,\,$, and
$\left(\hat{G}^{-1}_{02\downarrow\downarrow}\right)^\ast$, 
respectively, and 
subtract the resulting equations from one another to obtain,
\begin{equation}
\left[-i\frac{\partial}{\partial t} -
\frac{1}{m}\nabla_{\vec{r}}\nabla_{\vec{r}\,'}- H \right]
G^{-+}_{1\uparrow2\downarrow} =
-\int\left[\Sigma^{-\,\,a}_{1\uparrow3\alpha}
G^{a\,\,+}_{3\alpha2\downarrow} +
G^{-\,\,a}_{1\uparrow3\alpha} \Sigma^{a\,\,+}_{3\alpha2\downarrow}
\right]\,d^4X_3 \,\,,
\label{eq:kin1}
\end{equation}
where  $X$ and
$X^\prime$ are defined by
\begin{equation}
X=\frac{1}{2}(X_1+X_2)\,\,,\,\,\,\,\,X^\prime=X_1-X_2\,\,.
\end{equation}

We will be interested in departures from equilibrium which are (i) sufficiently
small so that the equations can be linearized and (ii) slowly varying
as functions of $X$ so that only leading terms in the gradient expansion
are required. The first approximation implies that
only one of the internal lines in any diagram contributing to 
$\Sigma^{ab}_{1\uparrow2\downarrow}$ or
$G^{ab}_{1\uparrow2\downarrow}$ may contain the transverse Green's
function $G^{ab}_{3\uparrow4\downarrow}\/$; by symmetry, 
the departures of the diagonal functions
$G^{ab}_{1\uparrow2\uparrow}\/$, $G^{ab}_{1\downarrow2\downarrow}\/$,
$\Sigma^{ab}_{1\uparrow2\uparrow}\/$ and 
$\Sigma^{ab}_{1\downarrow2\downarrow}\/$  from their equilibrium
form are already quadratic in  $G^{ab}_{3\uparrow4\downarrow}\/$
resulting in the decoupling of the transverse and longitudinal
Boltzmann equations in the linearized limit.
The second approximation requires that, in the rotating
frame, the characteristic scales
of variation of $G^{ab}_{3\uparrow4\downarrow}\/$ as a function of
$X$ should be large compared to all microscopic scales in the system:
\begin{equation}
\Delta L \gg \frac{1}{\Pd}\,,\,\,\,\,
\Delta t \gg \frac{m}{\Pd^2}\,,\,\,\,\,
\Delta L \gg \frac{1}{\Pu-\Pd}\,,\,\,\,\,
\Delta t \gg \frac{1}{ H}\,.
\label{eq:slow}
\end{equation}

The separation of scales discussed above is implemented as usual by
first transforming all quantities into the rotating frame and
then performing the gradient expansion in the Wigner transform.
The latter is defined as the Fourier transform with respect to
the relative coordinate, $X^\prime$ (here $F(X_1,X_2)$ is an arbitrary
function):
\begin{eqnarray}
F(X_1,X_2)&=&\int e^{-iP \cdot X^\prime} F(X,P) \frac{d^4P}{(2\pi)^4}\
,\,,
\,\,\,\,\,\,\nonumber \\ F(X,P)&=&\int
e^{iP\cdot
X^\prime}F(X+\frac{1}{2}X^\prime,X-\frac{1}{2}X^\prime)\,d^4
X^\prime\,,
\label{eq:wigner} \\
&&P=(\omega,\vec{p})\,\,,\,\,\,\,\,
P\cdot X^\prime=\omega t^\prime -\vec{p}\vec{r}\,'\,.\nonumber
\end{eqnarray}
In transforming into the rotating frame 
all transverse
self energies and Green's functions
acquire Larmor precession phase factors,
\begin{equation}
G^{ab}_{1\uparrow 2\downarrow}=e^{i H t}
\tilde{G}^{ab}_{1\uparrow2\downarrow}\,,\,\,\, 
\Sigma^{ab}_{1\uparrow 2\downarrow}=e^{i H t}
\tilde{\Sigma}^{ab}_{1\uparrow2\downarrow}\,.
\end{equation}

For example, for the mixed Fourier transform (only time arguments are 
shown) of a term 
${\cal K}$ containing 
$G^{ab}_{1\uparrow2\downarrow}$,
\[{\cal K}(t_1,t_2)=\int {\cal F}(t_1-t_3\,,t_3-t_4\,,t_4-t_2\,)
G^{ab}_{3\uparrow4\downarrow}\,dt_3dt_4\]
one obtains,
\begin{eqnarray}
&&{\cal K}(\omega, t)=e^{i H t}
\tilde{{\cal K}}(\omega, t)= \nonumber \\
&=&e^{i H t}\int {\cal F} 
\left(\omega - \frac{1}{2} H,
\omega - \omega^{\prime}, \omega +\frac{1}{2} H\right)
\Gt{ab}(\omega^{\prime},t)
\frac{d\omega^{\prime}}{2\pi}\,.
\label{eq:Tilde}
\end{eqnarray}
For the purposes of deriving the kinetic equations, the rotating 
frame function , $\tilde{{\cal K}}(\omega,t)$,
will then be treated as a slowly varying in the center-of-mass
time variable, $t$.

With the help of Eqns. (\ref{eq:wigner}--\ref{eq:Tilde}) 
the kinetic equation (\ref{eq:kin1}) can be rewritten as:
\begin{eqnarray}
&&\left(\frac{\partial}{\partial t} +
\frac{\vec{p}}{m}\frac{\partial}{\partial \vec{r}}\right) 
\left(-i\Gt{-+}(X,P)\right)= 
-\left[\Suu{-a}(P_-)\Gt{a+}(X,P)+\right. \nonumber \\
&&\left. +\St{-a}(X,P)\Gdd{a+}(P_+) +
\Guu{-a}(P_-) \St{a+}(X,P)+\Gt{-a}(X,P)\Sdd{a+}(P_+)\right]\,,
\label{eq:kin2}
\end{eqnarray}
where $P_\pm = (\omega \pm \frac{1}{2}H\,,\,\,\vec{p})$. Note that in Eqn. 
(\ref{eq:kin2}),
the dependence on $X$ only appears in the transverse
Green functions and self-energies as expected in the linearized limit.

The various self-energies
on the right hand side of 
Eqn. (\ref{eq:kin2}) are related to the
$T$-matrices as shown in Fig. \ref{fig:selfKeldysh}; since both ends of
the interaction line bear the same sign ($+$ or $-$), there are only
four kinds of the $T$-matrices involved ($T^{--}$, $T^{-+}$, $T^{+-}$
and $T^{++}$, see Fig. \ref{fig:selfKeldysh}).
All $T$-matrices are determined by Bethe-Salpeter equations
which, to lowest order in the transverse Green functions,
conserve the spin of each of the incoming particles. It is
convenient to rewrite the corresponding integral equations
by separating out the T-matrices involving
independent propagation on each of the Keldysh contours, 
$T^{--}_0$ and $T^{++}_0$;
the full $T^{ab}$-matrix is then obtained
by connecting factors of $T^{--}_0$ and $T^{++}_0$
by the appropriate $G^{-+}$ or $G^{+-}$ insertions
(see Fig. \ref{fig:TKeldysh}).
 
Thus $T^{--}_0$ and $T^{++}_0$ are determined by the Bethe-Salpeter
equations with no $G^{-+}$ or $G^{+-}$ intermediate lines.
It is easy
to see that, for $T=0$,
\begin{equation}
T^{--}_0(\vec{p}_1,\vec{p}_2;(s_0,\vec{s})) =
T(\vec{p}_1,\vec{p}_2;(s_0,\vec{s})) \,,\nonumber
\end{equation}
where $T(\vec{p}_1,\vec{p}_2;(s_0,\vec{s}))$ is the usual equilibrium
$T$-matrix calculated in Appendix \ref{app:self} (see Eqn.
(\ref{eq:Tequil}));
while at $T\neq 0$ we have:
\begin{eqnarray}
&&T^{--}_0(\vec{p}_1,\vec{p}_2;(s_0,\vec{s}))=
-\frac{4\pi a}{m} - \left(\frac{4\pi a}{m}\right)^2\times \nonumber \\
&&\times \int {\cal P} \left[ 
 \frac{{\cal N}^{\uparrow \downarrow}(\vec{k})}
{2s_0 -s^{\,2}/m+ 2
\mu - k^2/m}
-\frac{m}{p^2_2-k^2}
\right] \frac {d^3k}{(2\pi)^3}+ \label{eq:TT} \\
&&+i \pi \left(\frac{4\pi a}{m}\right)^2\int
\delta\left(2s_0-\frac{s^2}{m}+ 2\mu -\frac{k^2}{m} \right)\left(\Nud+ 2
n^\uparrow_{\vec{s}+\vec{k}}
n^\downarrow_{\vec{s}-\vec{k}}\right)\frac {d^3k}{(2\pi)^3} 
\nonumber
\end{eqnarray}
(see Equation (\ref{eq:Nud})).
Similarly, 
\begin{equation}
T^{++}_0(\vec{p}_1,\vec{p}_2;(s_0,\vec{s})) =
-\left(T^{--}_0(\vec{p}_1,\vec{p}_2;(s_0,\vec{s}))\right)^\ast \,.
\nonumber
\end{equation}

We note that 
the $(-+)$ insertions cannot modify the
physical scattering length\footnote{Formally 
this follows from the fact
that the integrals over energy and momentum in a ``bubble'' 
composed of two $(-+)$ 
Green's functions are convergent.} and thus, as can be seen from Fig. 
\ref{fig:TKeldysh} {\em b}, to second order in
$(N^{1/3}a)$, the T-matrices $T^{--}$ and $T^{++}$ are determined by the 
first term, 
\begin{eqnarray}
T^{--}(\vec{p}_1,\vec{p}_2;;(s_0,\vec{s}))& =&
T^{--}_0(\vec{p}_1,\vec{p}_2;(s_0,\vec{s})) \,,\nonumber \\
T^{++}(\vec{p}_1,\vec{p}_2;(s_0,\vec{s})) &=&
-\left(T^{--}_0(\vec{p}_1,\vec{p}_2;(s_0,\vec{s}))\right)^\ast \,.
\nonumber
\end{eqnarray}
Similarly, to second order in the scattering length, $T^{-+}$
is determined by the first diagram in Fig. \ref{fig:TKeldysh} {\em b},
proportional to
an integral of $T^{--} _0  G^{-+}_{\downarrow \downarrow} 
 G^{-+}_{\uparrow \uparrow}T^{++} _0$.

Finally, the self-energies required in the derivation of the
kinetic equations can now be calculated from Fig. \ref{fig:selfKeldysh}.
For example, the real and imaginary parts of the on-shell
self-energies (Fig. \ref{fig:selfKeldysh})
$\Suu{--}(\epsilon_\uparrow(\vec{p}),\vec{p})$ are given by Eqns. 
(\ref{eq:resigup}) and (\ref{eq:imsigup}) (see Eqn. (\ref{eq:TT})); 
similar equations hold for 
$\Sdd{--}(\epsilon_\uparrow(\vec{p}),\vec{p})$. Moreover,
\begin{equation}
\Suu{++}(P)=-\left(\Suu{--}(P)\right)^\ast
\,,\,\,\,\Sdd{++}(P)=-\left(\Sdd{--}(P)\right)^\ast\,.
\label{eq:sigrelt}
\end{equation}

We are now in the position to reexpress (\ref{eq:kin2}) as
a kinetic equation for
the transverse component of the density matrix (in the rotating
frame),
\begin{equation}
\Nt{p}=-i\int\Gt{-+}(X,P)\frac{d\omega}{2\pi}.
\nonumber
\end{equation}
We first note that the expansion of the off-diagonal self energies
$\Sigma^{-+}_{\alpha \beta}$ in powers of $N^{1/3}a$ starts with the
terms proportional to $(N^{1/3}a)^2$ (see Figs. \ref{fig:selfKeldysh} --
\ref{fig:TKeldysh}). In the diagonal self-energies,
$\Sigma^{--}_{\alpha \beta}$ and $\Sigma^{++}_{\alpha \beta}$, we
separate the first-order contributions,
\begin{eqnarray}
\Sigma^{(1)--}_{\uparrow \uparrow}(P)= \frac{4 \pi a} {m}
N_\downarrow\,\,&,& \,\,\,\,\,
\Sigma^{(1)++}_{\downarrow \downarrow}(P)= -\frac{4 \pi a} {m}
N_\uparrow\,\,,
\label{eq:sig1order} \\
\tilde{\Sigma}^{(1)--}_{\uparrow \downarrow}(X,P)=
-\tilde{\Sigma}^{(1)++}_{\uparrow \downarrow}(X,P)&=& 
\frac{4 \pi a} {m} \int n_{\uparrow\downarrow}(t,\vec{r},\vec{p}\,')
\frac{d^3p'}{(2\pi)^3}\,\,,
\nonumber
\end{eqnarray}
and move the corresponding terms  
in Eqn.  (\ref{eq:kin2}). to the l.\
h.\ s..  Expanding the Green's functions $G^{-+}_{\downarrow
\downarrow}$ and  $G^{-+}_{\uparrow \uparrow}$ with the help of
Dyson's equation (\ref{eq:sigdef1}) and Eqn. (\ref{eq:sig1order}), we
find, to linear order in \Nt{p}:
\begin{eqnarray}
\tilde{\Sigma}^{(1)--}_{\uparrow \downarrow}&&(X,P) G^{-+}_{\downarrow
\downarrow} (P_+)+G^{-+}_{\uparrow \uparrow}
(P_-)\tilde{\Sigma}^{(1)++}_{\uparrow \downarrow}(X,P)  = \\
&&= \frac{8
\pi^2 a }{m} i \delta(\omega - \epsilon(\vec{p}) + \mu)
(\Nu{p}-\Nd{p}) \int n_{\uparrow\downarrow}(t,\vec{r},\vec{p}\,')
\frac{d^3p'}{(2\pi)^3}+ {\cal O}(Na^3)\,\,.
\nonumber
\end{eqnarray}
Here we also used the following formulae for the non-interacting
longitudinal Green's functions:
\begin{eqnarray}
\Guu{(0)--}(P_-)&=&-\left(\Guu{(0)++}(P_-)\right)^{\ast}=
{\cal P}\frac{1}{\omega-\frac{p^2}{2m}+\mu}
+i \pi (2n^\uparrow_{\vec{p}}-1) \delta\left(\omega-\frac{p^2}{2m}
+\mu \right),\nonumber\\
\Guu{(0)+-}(P_-)&=&-2\pi i(1-\Nu{p})\delta(\omega - \frac{p^2}{2m} +
\mu ),
\label{eq:noninteq}\\
\Guu{(0)-+}(P_-)&=&2\pi i\Nu{p}\delta(\omega - \frac{p^2}{2m} +
\mu )\,,\nonumber
\end{eqnarray}
and similarly for the $\downarrow \downarrow$ components, in which case
$P_-$ is replaced by $P_+$.
Keeping the terms of up to the
second power of $a p_F$, we thus re-write Eqn.  (\ref{eq:kin2}) as
\begin{eqnarray}
&&\left(\frac{\partial}{\partial t} +
\frac{\vec{p}}{m}\frac{\partial}{\partial \vec{r}}\right) 
\left(-i\Gt{-+}(X,P)\right)+\frac{8
\pi^2 a }{m} i \delta(\omega - \epsilon(\vec{p}) + \mu)
(\Nu{p}-\Nd{p}) \int n_{\uparrow\downarrow}(t,\vec{r},\vec{p}\,')
\frac{d^3p'}{(2\pi)^3} + \nonumber\\
&&+\Suu{(1)--}(P_-)\Gt{-+}(X,P)+\Gt{-+}(X,P)\Sdd{(1)++}(P_+)= 
-\left[\Suu{(2)-a}(P_-)\Gt{(0)a+}(X,P)+\right.  
\label{eq:kin2a}\\
&&\left. +\St{(2)-a}(X,P)\Gdd{(0)a+}(P_+) +
\Guu{(0)-a}(P_-) \St{(2)a+}(X,P)+\Gt{(0)-a}(X,P)\Sdd{(2)a+}(P_+)\right]\,,
\nonumber
\end{eqnarray}
where to the leading order in interaction we can replace the
transverse Green's functions on the r.\ h.\ s. with
\begin{eqnarray}
\Gt{(0)-+}(X,P)&=&\Gt{(0)+-}(X,P)=\Gt{(0)++}(X,P)=
\Gt{(0)--}(X,P)= \nonumber \\
&& =2\pi i n_{\uparrow\downarrow}(t,\vec{r},\vec{p})
\delta(\omega - \frac{p^2}{2m}+\mu)\,.
\label{eq:Gt0}
\end{eqnarray}
Finally, using also  Eqns. (\ref{eq:sig1order}) and
(\ref{eq:noninteq}), we find upon integrating Eqn. (\ref{eq:kin2a})
over $\omega$:
\begin{eqnarray}
&&\left(\frac{\partial}{\partial t}+\frac{\vec{p}}{m}
\frac{\partial}{\partial \vec{r}} \right) \Nt{p} +
\frac{4\pi a}{m}i\int n_{\uparrow\downarrow}(t,\vec{r},\vec{p}\,')
\frac{d^3p'}{(2\pi)^3}(\Nu{p}-\Nd{p})- 
\nonumber \\
&&-\frac{4\pi a}{m}i(N_\uparrow-N_\downarrow)\Nt{p}
= -I_d[\Nt{p}]-I_e[\Nt{p}]\,.
\label{eq:kin3}
\end{eqnarray}
Above, $I_d[\Nt{p}]$ and $I_e[\Nt{p}]$ are 
the direct and exchange contributions to the
collision integral, respectively:
\begin{eqnarray}
I_d[\Nt{p}]&=&i\left\{\Suu{-+}(\Eu{p},\vec{p}) +
\Suu{(2)--}(\Eu{p},\vec{p}) + \right. \nonumber \\
&&+ \left. \Sdd{(2)++}(\Ed{p},\vec{p}) + \Sdd{-+}(\Ed{p},\vec{p})
\right\} \Nt{p}
\label{eq:direct}
\end{eqnarray}
and
\begin{eqnarray}
I_e[\Nt{p}]&=&i\St{-+}(X;\frac{p^2}{2m}-\mu,\vec{p})
(\Nu{p}+\Nd{p}-1) + \nonumber \\
&&+i\St{(2)--}(X;\frac{p^2}{2m}-\mu,\vec{p})\Nd{p}
+i\St{(2)++}(X;\frac{p^2}{2m}-\mu,\vec{p})\Nu{p}\,\,,
\label{eq:exchange}
\end{eqnarray}
where the superscript $(2)$ again implies that only
the quadratic terms in $a$ should be included in the
corresponding self-energies (we have used the notation (\ref{eq:energies})). 

Equations (\ref{eq:kin3}--\ref{eq:exchange}) are valid at any temperature; at
$T \rightarrow 0$, the kinetic equation can be simplified further by
assuming that 
\begin{equation}
\Nt{p}= 0\,,\,\,\,p<\Pd \,\,\,{\rm or}\,\,\,p>\Pu\,,
\label{eq:domain}
\end{equation}
This ansatz is consistent with the linearized equation:
to see this, first note that Eqns. (\ref{eq:noninteq}), (\ref{eq:Gt0}),
and (\ref{eq:domain}) together with
energy conservation
in the corresponding diagrams of  Figs. \ref{fig:selfKeldysh}, 
\ref{fig:TKeldysh} {\em b} imply 
\begin{eqnarray}
\St{-+}(\frac{p^2}{2m}-\mu, \vec{p})=&0\,, \,\,\,&p>\Pu \,, \nonumber \\
\St{+-}(\frac{p^2}{2m}-\mu, \vec{p})=&0\,, \,\,\,&p<\Pd\,.
\label{eq:transnodrag}
\end{eqnarray}
 From the transverse component of the self-energy identity
\cite{Volume10},
\begin{equation}
\Sigma^{++}_{1\alpha 2\beta}+\Sigma^{--}_{1\alpha 2\beta}+
\Sigma^{-+}_{1\alpha 2\beta}+\Sigma^{+-}_{1\alpha 2\beta}=0\,
\label{eq:selfidentity}
\end{equation}
it then follows that the exchange contribution to the
collision integral
(\ref{eq:exchange}) indeed vanishes outside of the intermediate
region, $p_{\downarrow} < p < p_{\uparrow}$ (the direct
term already vanishes by (\ref{eq:domain})).

Similarly, in the case of longitudinal self energies, 
energy conservation yields:
\begin{eqnarray}
\Suu{-+}(\epsilon_\uparrow(\vec{p}), \vec{p})=0\,,\,\,\,p>\Pu\,;\,\,\,&
\Sdd{-+}(\epsilon_\downarrow(\vec{p}), \vec{p})=0\,,&\,\,\,p>\Pd\,;
\nonumber \\ \Suu{+-}(\epsilon_\uparrow(\vec{p}),
\vec{p})=0\,,\,\,\,p<\Pu\,;\,\,\,& 
\Sdd{+-}(\epsilon_\downarrow(\vec{p}), \vec{p})=0\,,&\,\,\,p<\Pd\,
\nonumber
\end{eqnarray}
which, with the help of (\ref{eq:selfidentity}) and
(\ref{eq:sigrelt}) lead to:
\begin{eqnarray}
\Suu{-+}(\epsilon_\uparrow(\vec{p}),
\vec{p})&=&-2 i \I\Suu{}(\epsilon_\uparrow(\vec{p}), \vec{p})
\,,\,\,\,p<\Pu\,;\,\,\, \nonumber \\
\Sdd{-+}(\epsilon_\downarrow(\vec{p}), \vec{p})&=&
-2 i \I\Sdd{}(\epsilon_\downarrow(\vec{p}), \vec{p})\,,\,\,\,p<\Pd\,.
\label{eq:longnodrag}
\end{eqnarray}
Equations (\ref{eq:transnodrag}--\ref{eq:longnodrag}) then enable 
us to 
re-write the Boltzmann equation at $T \rightarrow 0$ in the 
form (\ref{eq:kin}). 

The finite temperature Boltzmann equation is derived by substituting 
the explicit expressions,
\begin{eqnarray}
&&\St{(2)++}\left(t,\vec{r};\frac{p^2}{2m}-\mu,\vec{p}\right)=
-\,\frac{(8\pi a)^2}{m}\int\left\{ {\cal P}
\frac{n^\uparrow_{\vec{s}+\vec{k}}+
n^\downarrow_{\vec{s}-\vec{k}}}{(\vec{p}-\vec{p}^{\,\prime})^2-4k^2} 
- \right. 
\label{eq:selftransverse} 
\\
&&\left. -\pi i \left[1-n^\uparrow_{\vec{s}+\vec{k}}
-n^\downarrow_{\vec{s}-\vec{k}}+2n^\uparrow_{\vec{s}+\vec{k}}
n^\downarrow_{\vec{s}-\vec{k}}\,
\right] \delta\left((\vec{p}-\vec{p}^{\,\prime})^2-4k^2\right)
\right\}n_{\uparrow\downarrow}(t,\vec{r},\vec{p}^{\,\prime})
\frac{d^3k\,d^3p^\prime}{(2\pi)^6}\,,\nonumber
\end{eqnarray}
\begin{eqnarray}
\Suu{-+}(\Eu{p},\vec{p}) & = & 2 \pi i \frac{(4 \pi a)^2}{m} \int
n^\downarrow_{\vec{s}-\vec{k}}n^\uparrow_{\vec{s}+\vec{k}}
\left(n^\downarrow_{\vec{p}\,'} - 1\right)
\delta \left( \frac{1}{4}(\vec{p} - \vec{p}^{\,\prime})^2-k^2 \right)
\frac{d^3k\,d^3p^\prime}{(2 \pi)^6} \,,\nonumber\\
\Sdd{-+}(\Ed{p},\vec{p}) & = & 2 \pi i \frac{(4 \pi a)^2}{m} \int
n^\downarrow_{\vec{s}-\vec{k}}n^\uparrow_{\vec{s}+\vec{k}}
\left(n^\uparrow_{\vec{p}\,'} - 1\right)
\delta \left( \frac{1}{4}(\vec{p} - \vec{p}^{\,\prime})^2-k^2 \right)
\frac{d^3k\,d^3p^\prime}{(2 \pi)^6} \, \nonumber
\end{eqnarray}
into (\ref{eq:kin3}--\ref{eq:exchange}) 
(here and below $\vec{s}=(\vec{p}+\vec{p}^{\,\prime})/2$)
to obtain the equation valid at any temperature and magnetic field
in the dilute s-wave limit:
\begin{eqnarray}
&&\left(\frac{\partial}{\partial t}
+\frac{\vec{p}}{m}\frac{\partial}{\partial \vec{r}}\right) \Nt{p} =
\Nt{p}\,   \times \nonumber \\
&&\times \int \left\{ i \left[\frac{4 \pi a}{m} -
B(\vec{p},\vec{p}^{\,\prime})\right] (n^\uparrow_{\vec{p}^{\,\prime}}-
n^\downarrow_{\vec{p}^{\,\prime}})-
 A(\vec{p},\vec{p}^{\,\prime})\right\}
\frac{ d^3 p^\prime}{(2 \pi)^3}- \nonumber \\
&&- \int\left\{ i\left[\frac{4 \pi a}{m}-
B(\vec{p}^{\,\prime},\vec{p})\right] (n^\uparrow_{\vec{p}}-
n^\downarrow_{\vec{p}})-
 A(\vec{p}^{\,\prime},\vec{p})\right\}
n_{\uparrow\downarrow}(t,\vec{r},\vec{p}^{\,\prime})
\frac{ d^3 p^\prime}{(2 \pi)^3} \,\,.
\label{eq:Diffkin}
\end{eqnarray}
Here 
\begin{equation}
A(\vec{p},\vec{p}^{\,\prime})  =  \frac {\pi (4 \pi a)^2}{m} \int 
\left[ 2 n^\uparrow_{\vec{s} + \vec{k}} n^\downarrow_{\vec{s} -
\vec{k}} + (n^\uparrow_{\vec{p}^{\,\prime}} +
n^\downarrow_{\vec{p}^{\,\prime}}) \Nud \right]  
\delta \left(
\frac{1}{4} (\vec{p} - \vec{p}^{\,\prime})^2 - k^2 \right) \frac{d^3
k} {(2 \pi )^3}\,
\label{eq:relgeneral}
\end{equation}
and
\begin{equation}
B(\vec{p},\vec{p}^{\,\prime})=\frac{(8\pi
a)^2}{m} \int {\cal P} \frac
{n^\uparrow_{\vec{s}+\vec{k}} + n^\downarrow_{\vec{s}-\vec{k}}}
{(\vec{p}- \vec{p}^{\,\prime})^2-4k^2} \frac {d^3 k}{(2\pi)^3}\,\,.
\label{eq:srgeneral}
\end{equation}
We note that Eqns. (\ref{eq:Diffkin}--\ref{eq:srgeneral}) are in
complete agreement
with the appropriate limit (see inequalities (\ref{eq:slow}) above)
of the kinetic equation of
Ref. \cite{Mullin88}\footnote{The origins of the difference in
numerical factors between Eqns. (\ref{eq:Diffkin}--\ref{eq:srgeneral})
and the results of Ref. \cite{Sergey} (which contains another
derivation of Boltzmann equation within Keldysh formalism) are unknown
to us.}. 
Upon integration of Eqn. (\ref{eq:Diffkin}) over the momentum space 
the two terms on the r.\
h.\ s. (``direct'' and ``exchange'' terms) cancel each other,
and, according to Eqns. (\ref{eq:defM}--\ref{eq:defJ}), we are left as expected
with the continuity equation (\ref{eq:macromagT0}). 

Elementary integration in Eqn. (\ref{eq:relgeneral}) leads to the
following expression for 
$A(\vec{p},\vec{p}^{\,\prime})$: 
\begin{eqnarray}
&&\frac{A(\vec{p},\vec{p}^{\,\prime})}{2 \pi a^2 T}  = 
\frac{2}{s}\cdot \frac{1}{1-{\rm exp}(2w)} 
\left[ \ln \frac{1+{\rm exp} (s u - w+h)}{1+{\rm
exp} (-s u -w +h)} 
- \right. \nonumber \\
&& \left. - \ln\frac{1+{\rm exp} (s u + w + h)}{1+{\rm exp} 
(-s u + w+h)} \right]+\label{eq:appprime}\\ 
&&+ (n^\uparrow_{\vec{p}^{\,\prime}}+
n^\downarrow_{\vec{p}^{\,\prime}}) 
 \left( 2u - \frac{1}{s} \left[ \ln \frac{1+ {\rm exp}
 (s u - w+h) }{1+{\rm exp} (-s u - w+h )} + 
\ln \frac{1+{\rm exp} 
(s u - w-h)}{1+{\rm exp} 
(-s u -w-h) } \right] \right) \,, 
\nonumber
\end{eqnarray} 
where we use the following notation:
\[ u = \frac{\mid\vec{p} - \vec{p}^{\,\prime}\mid}{2mT}\,\,\,\,\,,
\,\,\,\,\,\,\,
w=\frac{(p^2+{p^\prime}^2)-4m\mu}{4mT}\,,\,\,\,\,\,\,h=\frac{H}{2T} \,. \]

In the limit of low
temperatures and small polarizations, when
$T, H \ll \mu$,
only the electrons from the vicinity of the Fermi spheres may
participate in the scattering. Therefore, in the expression
(\ref{eq:appprime}) we may assume that
$p \approx p^\prime \approx (2 m \mu)^{1/2}$ and 
$w \stackrel{<}{\sim} {\rm max}(T, H)/T$.
Moreover, it is easy to check that the contribution of backward and forward
scattering regions (where the product $su$ may be arbitrarily small) is
negligible, so one can also assume that $s u T \gg {\rm max}(T, H)$.

With this assumptions, we obtain  Eqn. (\ref{eq:aaproxT}) for
$A(\vec{p},\vec{p}^{\,\prime})$. The quantity
$B(\vec{p},\vec{p}^{\,\prime})$ at $T, H \ll \mu$ is given by
Eqn. (\ref{eq:baproxT}).

\figure{The upper and lower bounds of the relaxation time
functional $\tau_\bot$ as  functions of $d=\Pd/\Pu$. Solid lines
correspond to the functions $a^2\Pu^4\tau_\bot^\pm(d)/m\,$. The dashed
line represents the diffusion time $\tau_D$ (see Eqn. (\ref{eq:diftimeT0})).
\label{fig:bounds}}

\figure{Spin diffusion coefficient in the high-field region
$H \stackrel{>}{\sim} T$. The transverse
spin diffusion coefficient $D_\bot$ (Eqn. (\ref{eq:diffhigh})), 
the longitudinal coefficient $D_\parallel$,
and the simple fit \cite{Candela,Karen3}
$\left(1/D_\bot^{(0)}+1/D_\parallel\right)^{-1}$  (Eqn.
(\ref{eq:difffit})) 
for $D_\bot$ ($D_\bot^{(0)}=3\pi v_F^2/8ma^2H^2$ is the limiting
value of $D_\bot$ at $T \rightarrow 0$) are represented by
the solid, dashed and dotted lines, respectively.
\label{fig:diffhigh}}

\figure{The crossover at $T \sim T_c=\sqrt{H\epsilon_F/ap_F}$ (i.\ e.\
$\zeta \sim 1$). Fig. \ref{fig:diffcross} {\em a} shows the ratio
$D_\bot/D_\parallel$. The spin-rotation parameter $\xi$ is plotted in
Fig. \ref{fig:diffcross} {\em b}.
\label{fig:diffcross}}

\figure{The functions $f(p)$ that solve the steady-state equation
(\ref{eq:difmicro}) in the isotropic limit $\zeta \ll 1$ (solid line)
and in the intermediate regime, $\zeta \gg 1$ and $H \ll T$
(dashed line) at the same temperature $T$.
\label{fig:changeform}}

\figure{The low-field behaviour of the spin-rotation parameter $\xi$
in the concentrated $^3{\rm He}$--$^4{\rm He}$ mixture. The points
represent the experimental data of Reference~\cite{Owers}, the dashed
line -- the $T^{-2}$ fit for the behaviour of $\xi$ below the crossover,
and the solid line is our theoretical result for the $s$-wave
approximation (with the values of $T$ and $\xi$ scaled by appropriate
factors). Note, that the latter is {\em not} a straight line. 
\label{fig:spinrot}}

\figure{The diagrams for the self-energies
$\Suu{}(\vec{p})$ and  $\Sdd{}(\vec{p})$.
\label{fig:self}}

\figure{ The $T$-matrix $T^{ab}(\vec{p}_1,\vec{p}_2;(s_0,\vec{s}))$ and
the diagrams for 
the self energies $\Suu{ab}$, 
$\Sdd{ab}$ and 
$\Sigma_{\uparrow\downarrow}^{ab}$ in the Keldysh technique; 
the indices $a$ and $b$ may take
the values $+$ or $-$.
\label{fig:selfKeldysh}}

\figure{({\em a}). Bethe -- Salpeter equations for the auxiliary
$T$-matrices  $T^{--}_0$ and $T^{++}_0$.
({\em b}). The diagrammatic series for the $T$-matrices $T^{--}$
and $T^{-+}$. The series for $T^{++}$ and $T^{+-}$ can be drawn in
a similar way. Within the second order in
$N^{1/3}a$ only the first terms in these diagrammatic series should
actually be taken into account.
\label{fig:TKeldysh}}


\begin{references}
\bibitem{bounds}The variational bounds were obtained by varying
the relaxation time functional $\tau_\bot[W(p)]$ (defined as in 
Eqn. (\ref{eq:deftau}) of Sect. \ref{sec:smallpol}),
over real functions $W(p)$ such that $W(p)>0$ for $p_\downarrow < p 
< p_\uparrow$ and $W(p) \equiv 0$ otherwise.
\bibitem{misprint}In Equation (\ref{eq:xihigh}) we
have corrected the misprint of Ref. \cite{We} (equation (10)).
\end{references}
\end{document}